# Case-deletion importance sampling estimators: Central limit theorems and related results

**Ilenia Epifani**

*Department of Mathematics*
*Politecnico di Milano*
*I-20133 Milano, Italy*
*e-mail:* ilenia.epifani@polimi.it
*url:* www1.mate.polimi.it/∼ileepi/

**Steven N. MacEachern**[†] **and Mario Peruggia**[‡]

*Department of Statistics*
*The Ohio State University*
*Columbus, OH 43210-1247, USA*
*e-mail:* snm@stat.osu.edu; peruggia@stat.osu.edu
*url:* www.stat.osu.edu/∼snm/; www.stat.osu.edu/∼peruggia/

**Abstract:** Case-deleted analysis is a popular method for evaluating the influence of a subset of cases on inference. The use of Monte Carlo estimation strategies in complicated Bayesian settings leads naturally to the use of importance sampling techniques to assess the divergence between full-data and case-deleted posteriors and to provide estimates under the case-deleted posteriors. However, the dependability of the importance sampling estimators depends critically on the variability of the case-deleted weights. We provide theoretical results concerning the assessment of the dependability of case-deleted importance sampling estimators in several Bayesian models. In particular, these results allow us to establish whether or not the estimators satisfy a central limit theorem. Because the conditions we derive are of a simple analytical nature, the assessment of the dependability of the estimators can be verified routinely before estimation is performed. We illustrate the use of the results in several examples.

[*]The author thanks the Department of Statistics at the Ohio State University for the kind hospitality during her visit
[†]Supported in part by NSF Awards No. DMS-0072526, SES-0437251, and DMS-0605041.
[‡]Supported in part by NSF Awards No. SES-0214574, SES-0437251 and DMS-0605052.

## 1. Introduction

Complex Bayesian models are fit with simulation techniques. A Monte Carlo method is used to generate a sample from the posterior distribution, and this sample is used to estimate many quantities, such as posterior means and variances of parameters, posterior probabilities of events, predictive distributions of future cases, etc. For a complete analysis, one examines the data, looking for outliers and influential cases. One also considers information external to the model which suggests groups of cases that may depart from the model. When interesting groups of cases are found, they are dropped from the data set, and estimates are recomputed. The resulting case-deleted posterior distribution and the case-deleted estimates are of interest, as are the changes in the posterior and estimates. Substantial changes in posterior or estimates may lead to refinement of the model. Cross-validation also relies on case-deletion, as formalized by the conditional predictive ordinate (CPO) (see, for example, p. 47 and p. 284 of [4]).

Case-deleted posterior distributions are examined through importance sampling. The large sample from the full posterior distribution is reweighted, as suggested for example in [21] and [23], to compute summaries with respect to the case-deleted posterior distribution. Examples of this and similar approaches are presented in [3, 15, 16, 25, 26] and [27]. As shown in [13] it is essential for the importance sampling weights to have finite variance. If the 2nd moment of the weights does not exist, typical estimators will not follow a $n^{1/2}$ asymptotic, nor will they follow a central limit theorem.

It is shown in [19] that, for the case of a popular Bayesian linear model with conjugate priors, whether or not the weight function for a single case-deletion has finite 2nd moment depends on simple conditions involving the scale parameter of the prior distribution of the error variance, the leverage of the observation being deleted, its residual, and the total residual sum of squares. In this article, we expand upon the results of [19] in several directions. We first analyze the situation of multiple case-deletions and provide necessary and sufficient conditions for the $r$th ($r > 1$) posterior moment of the weight function to be finite. This allows us to treat a group of observations coherently, thereby capturing synergistic effects of similar cases. We extend the results to much broader classes of prior distributions, so that we can handle nonconjugate as well as conjugate priors. This is accomplished by formally defining classes of distributions that are thick or thin tailed with respect to the conjugate priors. This extension is coupled with two devices, bounding functions and adjustment of the prior, to allow us to establish a connection between a finite $r$th moment of the weight function and the finiteness of the 2nd moment for a variety of functions. The existence of two moments for these functions implies that a central limit theorem holds for an estimator. As in [19], the conditions are on sample size, leverage and an adjusted residual sum of squares.

In addition to the linear model, we provide results for the Michaelis-Menton (MM) model. The MM model is nonlinear, but has the property that, conditional on one parameter, the mean structure is linear in the remaining parameters. Making use of conditional linearity, we develop uniform versions of the conditions for the linear model that ensure existence of the weight function's $r$th moment in the MM model. Many other models are conditionally linear (among them linear regression used in conjunction with Box-Cox transformations or linear regression along with Box-Tidwell transformations). We pursue further extensions of the linear model, deriving results for the logistic regression model.

Our results have a very practical implication. They let us determine, quickly and analytically, whether central limit theorems hold for particular functionals. If central limit theorems hold, then we can pursue the strategy of fitting the model to the full data set and using importance sampling to estimate the functionals under case-deleted posteriors. If central limit theorems do not hold, we must alter our inferential strategy, either using more sophisticated importance sampling techniques (such as the importance link function technique introduced in [17]) or fitting the model for particular case-deleted data sets with separate Monte Carlo simulations.

By providing conditions under which $r$ moments of the case-deleted weight function exist, our theorems go beyond the typical central limit theorem results that rely on the existence of second moments. This is important for two reasons. First, one may be interested in functionals where higher order moments of the case-deleted weight function come into play (see, for example, estimation of $\chi^2$ divergence in [26] and [27]). Second, the number of moments which exist for the deletion of particular cases can be used as a measure of their influence, thus allowing one to asses influence along a continuum. The connection between influence and moment conditions is elucidated by applying results presented in [7] and [8], which, for an arbitrary, non-negative random variable $X$, contain the definition of a quantity called the *moment index* of $X$. Denoting by $W$ the weight function resulting from the deletion of a given set of cases, its moment index $r^*$ is the least upper bound on the number of moments which exist. This represents a quantitative summary of the limiting tail behavior of the case-deleted weight function in the sense that, as stated in [7] and [8], $r^* = \liminf_{t\to\infty}[\log P(W > t)]/[\log(1/t)]$. A larger moment index corresponds to a larger class of functions for which the central limit theorem exists. Practical illustration of these ideas are presented in Sections 4 and 6.

This article is laid out as follows. Section 2 contains preliminary results and formal definitions of thick and thin tails. Section 3 provides conditions for the (non)existence of the $r$th moment of the case deletion weight function in the linear model. Section 4 gives conditions on moments' existence for the MM model, and Section 5 gives parallel results for the logistic regression model. Section 6 shows how the results can be used to establish central limit theorems. A summary of sufficient conditions on the weight function's moments to ensure a central limit theorem for several popular Bayesian measures of influence are presented in Table 2. The section also shows the results in action, investigating both measures of influence and their impact on model development in a multiple linear regression setting. The final section contains concluding remarks. Technical details of proofs are left to the appendix.

## 2. Notation and preliminary results

Each Bayesian model considered in this article depends on a finite dimensional parameter vector $s = (s_1, \ldots, s_k)$. Suppose that a set of observations $\boldsymbol{y} = (y_1, \ldots, y_n)$ is collected and let $p(s) = p(s|\boldsymbol{y})$ denote the full posterior density for $s$. Let $\mathcal{I}$ denote the set of indices to be deleted from the analysis and let $I$ be its cardinality. Let $\boldsymbol{y}_{\setminus\mathcal{I}}$ represent the $n - I$ observations remaining after the indices in $\mathcal{I}$ are omitted with $p_{\setminus\mathcal{I}}(s) = p_{\setminus\mathcal{I}}(s|\boldsymbol{y}_{\setminus\mathcal{I}})$ denoting the corresponding case-deleted posterior density. Furthermore, let $q(s) = q(s, \boldsymbol{y})$ and $q_{\setminus\mathcal{I}}(s) = q(s, \boldsymbol{y}_{\setminus\mathcal{I}})$ denote functions computable at every point $(s, \boldsymbol{y})$ and proportional to the joint prior densities (e.g., prior $\times$ data likelihood) of $(s, \boldsymbol{y})$ and $(s, \boldsymbol{y}_{\setminus\mathcal{I}})$, respectively.

Suppose that a sample $z_1, \ldots, z_M$ from $p(s)$ is available. In a typical application this will be either an independent sample or a dependent sample from an ergodic Markov chain. We wish to construct an estimate of $\mathrm{E}_{p_{\setminus\mathcal{I}}}[g(s)] = \int g(s)p_{\setminus\mathcal{I}}(s)\, d(s)$, for some real valued function $g(s)$ such that $\int |g(s)|\, p_{\setminus\mathcal{I}}(s)\, ds < \infty$. This can be done by computing a Monte Carlo sum in which the individual elements $g(z_m)$ are reweighted. Typically, $p(s)$ and $p_{\setminus\mathcal{I}}(s)$ are not available because their normalizing constants are unknown and only $q(s)$ and $q_{\setminus\mathcal{I}}(s)$ are directly computable. In that case we can define the weight function $w_{\setminus\mathcal{I}}(s) = q_{\setminus\mathcal{I}}(s)/q(s)$ and estimate the expectation by:

$$\hat{\mathrm{E}}_{p_{\setminus\mathcal{I}}}[g(s)] = \left(\sum_{m=1}^{M} w_{\setminus\mathcal{I}}(z_m)g(z_m)\right) \Bigg/ \left(\sum_{m=1}^{M} w_{\setminus\mathcal{I}}(z_m)\right). \tag{2.1}$$

The denominator in Equation (2.1) divided by $M$ estimates the ratio of the two unknown normalizing constants. Thus, if $p(s)$ and $p_{\setminus\mathcal{I}}(s)$ are available, $w_{\setminus\mathcal{I}}(s)$ can be replaced by $w^*_{\setminus\mathcal{I}}(s) = p_{\setminus\mathcal{I}}(s)/p(s)$ in the numerator and the denominator can be replaced by $M$, resulting in the related estimator that we denote by $\hat{\mathrm{E}}^*_{p_{\setminus\mathcal{I}}}[g(s)]$. In both cases, the resulting estimators are consistent under mild assumptions (see [13] for the case of i.i.d. samples and [24] for the case of samples from ergodic Markov chains). Throughout the article we refer to estimators of the form $\hat{\mathrm{E}}_{p_{\setminus\mathcal{I}}}[g(s)]$ and $\hat{\mathrm{E}}^*_{p_{\setminus\mathcal{I}}}[g(s)]$ as *case-deleted importance sampling estimators*.

The prior distribution plays a large role in determining whether the Estimator (2.1) is asymptotically normal. To ensure asymptotic normality for $\hat{\mathrm{E}}_{p_{\setminus\mathcal{I}}}[g(s)]$, we need both $\int w^2_{\setminus\mathcal{I}}(s)g^2(s)p(s)\, ds < \infty$ and $\int w^2_{\setminus\mathcal{I}}(s)p(s)\, ds < \infty$. Finiteness of these integrals is unchanged by substitution of $w^{*2}_{\setminus\mathcal{I}}(s)$ for $w^2_{\setminus\mathcal{I}}(s)$. (See Section 6 for further discussion of conditions for the asymptotic normality of both $\hat{\mathrm{E}}_{p_{\setminus\mathcal{I}}}[g(s)]$ and $\hat{\mathrm{E}}^*_{p_{\setminus\mathcal{I}}}[g(s)]$.) In many instances, a prior distribution with sharp enough tails will ensure that these integrals are finite while a flatter tailed prior will lead to infinite integrals.

The upcoming lemma enables us to work easily with priors having different tails. In particular, it enables us to derive preliminary results for conjugate prior distributions, and then to quickly extend the results to non-conjugate prior distributions. Use of the lemma is demonstrated in the examples.

To set up the lemma, we first define the basic notation. Let

$$S_i = \int \frac{f(x)\pi_i(x)}{\int f(u)\pi_i(u)du}\, h(x)\, dx = c_i^{-1} \int f(x)\pi_i(x)h(x)\, dx,$$

for $i = 0, 1$. The functions $f, \pi_i$ and $h$ are assumed to be non-negative. The constants $c_i$ are assumed to be finite and positive. Let $0 < b < B < \infty$.

**Lemma 2.1.** *If, for all $x$, $\pi_0(x)/\pi_1(x) < B$, then $S_0 = \infty$ implies $S_1 = \infty$. If, for all $x$, $b < \pi_0(x)/\pi_1(x)$, then $S_0 < \infty$ implies $S_1 < \infty$. If, for all $x$, $b < \pi_0(x)/\pi_1(x) < B$, then $S_0 < \infty$ if and only if $S_1 < \infty$.*

A device that we have found useful is a formal description of thinner and thicker tailed distributions. Since the prior distributions that we consider here are all absolutely continuous with respect to Lebesgue measure on $\mathcal{R}^k$, we use a simple definition that suffices for our purposes. We describe the result in terms of a distribution for a parameter since that is how we will use the result.

Consider a parameter $s \in S$. The parameter space $S$ is taken to be $\mathcal{R}^k$. Let $\mathcal{F}$ represent a set of distributions on $s$, all of which have densities with respect to Lebesgue measure. The following definition concerns the relationship between another distribution, $g$, and the set of distributions $\mathcal{F}$.

**Definition 2.1.** The density $g$ is said to be *thick-tailed* with respect to $\mathcal{F}$ if, for each $f \in \mathcal{F}$ and for each sequence $s_t$ with $||s_t|| \to \infty$ as $t \to \infty$, $\lim_{t\to\infty} g(s_t)/f(s_t) = \infty$.

**Definition 2.2.** The density $g$ is said to be *thin-tailed* with respect to $\mathcal{F}$ if, for each $f \in \mathcal{F}$ and for each sequence $s_t$ with $||s_t|| \to \infty$ as $t \to \infty$, $\lim_{t\to\infty} g(s_t)/f(s_t) = 0$.

We note that these definitions capture the general notion of which distributions are thicker or thinner tailed than others. For example, a $t$ distribution will be thicker tailed than the class of normal distributions. A one-dimensional normal distribution will be thinner tailed than the Laplace distribution. A $t$ distribution with 5 degrees of freedom will be thicker tailed than a $t$ distribution with 7 degrees of freedom, etc. We also note that a normal distribution with variance $\sigma^2$ is thicker tailed than a normal distribution with variance $c\sigma^2$ if $c < 1$.

## 3. A Bayesian linear model

In [19] the author considers a standard specification of the Bayesian linear model and derives necessary and sufficient conditions for the variance of the case-deleted importance sampling weight function to be finite when a single observation is omitted. Loosely, the conditions for a finite variance stated in [19] can be described as (a) small leverage for the deleted case, (b) large enough sample size, and (c) small enough residual for the deleted case. In this section we extend the results of [19] in two different directions: we analyze the situation of multiple case-deletions and provide necessary and sufficient conditions for the $r$th ($r > 1$) posterior moment of the case-deleted weight function to be finite. Our conditions are also on leverage, sample size and residual. In addition, we extend the results to nonconjugate models by considering the tail behavior of the prior distribution. In Section 6, these results are used to establish central limit theorems for a broad class of importance sampling estimators.

Let the $n \times 1$ vector of observations $\boldsymbol{Y}$ be distributed as

$$\boldsymbol{Y}|\boldsymbol{\theta}, \sigma^2 \sim N\left(X\boldsymbol{\theta}, \sigma^2\mathbf{I}\right), \tag{3.1}$$

where $\mathbf{I}$ denotes the identity matrix and $X$ denotes an $n \times k$ design matrix of rank $k$. Assume that the variance $\sigma^2$, having an inverse gamma prior distribution with known positive parameters $\alpha$ and $\beta$, is independent of the $k \times 1$ vector of regression parameters $\boldsymbol{\theta} = (\theta_0, \ldots, \theta_{k-1})^T$ having a proper prior density $\pi_1$ with full support $\mathcal{R}^k$, i.e.,

$$\boldsymbol{\theta} \sim \pi_1 \quad \perp \quad \sigma^2 \sim IG\left(\alpha, \beta\right). \tag{3.2}$$

To describe conditions under which moments of the case-deleted weight function exist, we introduce some additional quantities. Let $H = X(X^TX)^{-1}X^T$ and $\mathrm{RSS} = \boldsymbol{y}^T(\mathbf{I} - H)\boldsymbol{y}$ denote the projection matrix and the residual sum of squares from the least squares fit of the full data set, respectively. The index set, $\mathcal{I}$, consists of the indices of the $I$ cases to be deleted. Given the index set $\mathcal{I}$, let $\boldsymbol{Y}_{\mathcal{I}}$ be the $I \times 1$ random vector of observations $Y_i$, with $i \in \mathcal{I}$, and let $X_{\mathcal{I}}^T$ be the $I \times k$ submatrix of the $I$ rows of $X$ indexed by $\mathcal{I}$. Define the leverage of set $\mathcal{I}$ to be the principal minor of $H$ corresponding to $\mathcal{I}$: $H_{\mathcal{I}} = X_{\mathcal{I}}^T(X^TX)^{-1}X_{\mathcal{I}}$, and define $\boldsymbol{e}_{\mathcal{I}}$ to be the $I \times 1$ vector of the elements indexed by $\mathcal{I}$ in the vector of the ordinary residuals $\boldsymbol{e} = (\mathbf{I} - H)\boldsymbol{y}$, i.e., $\boldsymbol{e}_{\mathcal{I}} = \boldsymbol{y}_{\mathcal{I}} - X_{\mathcal{I}}^T(X^TX)^{-1}X^T\boldsymbol{y}$. Finally, for each $r > 0$, if the $I \times I$ matrix $(\mathbf{I} - r\,H_{\mathcal{I}})$ is non-singular, let $\mathrm{RSS}^*_{\setminus\mathcal{I}}(r) = \mathrm{RSS} - r\,\boldsymbol{e}_{\mathcal{I}}^T(\mathbf{I} - r\,H_{\mathcal{I}})^{-1}\boldsymbol{e}_{\mathcal{I}}$. When $I = 1$, so that $\mathcal{I} = \{i\}$, $H_{\mathcal{I}} = \boldsymbol{x}_i^T(X^TX)^{-1}\boldsymbol{x}_i$ is the leverage of $i$th observation, say $h_{ii}$, $e_i = y_i - \boldsymbol{x}_i^T(X^TX)^{-1}X^T\boldsymbol{y}$ is the residual of observation $i$, and $\mathrm{RSS}^*_{\setminus i}(r) = \mathrm{RSS} - r\,e_i^2/(1 - rh_{ii})$. When $r = 1$, $\mathrm{RSS}^*_{\setminus\mathcal{I}}(r)$ is the residual sum of squares from the least squares fit of the case-deleted data set. Letting $s = (\boldsymbol{\theta}, \sigma^2)$, the unnormalized importance sampling weight function resulting from the deletion of the $I$ cases indexed by $\mathcal{I}$ is given by

$$w_{\setminus\mathcal{I}}(s) = (\sigma^2)^{I/2}\exp\left\{1/(2\sigma^2)(\boldsymbol{Y}_{\mathcal{I}} - X_{\mathcal{I}}^T\boldsymbol{\theta})^T(\boldsymbol{Y}_{\mathcal{I}} - X_{\mathcal{I}}^T\boldsymbol{\theta})\right\}. \tag{3.3}$$

This functional form of the weight results from ignoring normalizing constants not depending on the model parameters and from canceling the common factors in the numerator and the denominator represented by the prior and by the portion of the Gaussian likelihood which corresponds to the undeleted cases.

For the Bayesian linear model specified by Equations (3.1) and (3.2) the following theorem holds.

**Theorem 3.1.** *Let* $\boldsymbol{Y} \mid \boldsymbol{\theta}, \sigma^2 \sim N(X\boldsymbol{\theta}, \sigma^2\mathbf{I})$. *Let* $\lambda_1 \leq \cdots \leq \lambda_I$ *denote the eigenvalues of* $H_{\mathcal{I}}$ *and assume that* $\lambda_i \neq 1/r$, *for all* $i = 1, \ldots, I$.
(*i*) *If the prior distribution follows specification* (3.2), *then the case-deleted weight function* $w_{\setminus \mathcal{I}}(s)$ *has a finite* $r$th *moment with respect to the full posterior* $p(s)$ *if*

$$(a)\ \lambda_I < 1/r \quad and \quad (b)\ n/2 + \alpha > r\,I/2 \quad and \quad (c)\ \mathrm{RSS}^*_{\setminus \mathcal{I}}(r) > -2/\beta.$$

*Conversely, the* $r$th *moment of* $w_{\setminus \mathcal{I}}(s)$ *is infinite if*

$$(a')\ \lambda_I > 1/r \quad or \quad (b')\ n/2 + \alpha \leq r\,I/2 \quad or \quad (c')\ \mathrm{RSS}^*_{\setminus \mathcal{I}}(r) < -2/\beta.$$

(*ii*) *If the noninformative prior* $\pi(\boldsymbol{\theta}, \sigma^2) \sim 1/\sigma^2$ *is used, then conditions* (*a*) *and* (*a*′) *remain unchanged, and conditions* (*b*), (*c*), (*b*′) *and* (*c*′) *become:* (*b*) $n > rI + k$, (*b*′) $n \leq rI + k$, (*c*) $\mathrm{RSS}^*_{\setminus \mathcal{I}}(r) > 0$ *and* (*c*′) $\mathrm{RSS}^*_{\setminus \mathcal{I}}(r) < 0$.

*Remark* 3.1. Theorem 3.1 includes the problem investigated in [19] as a special case. There, the author takes $r = 2$ and specifies the prior distribution on $(\boldsymbol{\theta}, \sigma^2)$ as $\boldsymbol{\theta} \mid \Sigma \sim N(\boldsymbol{\theta}_0, \Sigma)$, $\sigma^2 \sim IG(\alpha, \beta)$ and $\Sigma \sim IW(\nu R, \nu)$, with conditional independence at all stages of the model. The parameter $\boldsymbol{\theta}_0 \in \mathcal{R}^k$ is a known mean vector, $\alpha$ and $\beta$ are known positive constants, and $IW(\nu R, \nu)$ is an inverse Wishart distribution with $\nu$ a known integer greater than or equal to $k$ and $R$ a known $k \times k$ positive definite matrix.

*Remark* 3.2. The statement of Theorem 3.1 involves the eigenvalues of the $I \times I$ matrix $H_{\mathcal{I}}$. In typical applications, the cardinality $I$ of the set of observations being deleted will be fairly small and the calculation of the eigenvalues can be accomplished quickly with standard software. For the illustrative examples presented in the article, we computed all eigenvalues using the R function `eigen()`.

Theorem 3.1 holds for any proper prior distribution on $\boldsymbol{\theta}$ having full support on $\mathcal{R}^k$, provided the parameters $\boldsymbol{\theta}$ and $\sigma^2$ are independent and the prior for $\sigma^2$ is $IG(\alpha, \beta)$. This follows from the form of the likelihood function, which, for fixed $\sigma^2$, is an exponential function with quadratic argument in $\boldsymbol{\theta}$, and, for fixed $\boldsymbol{\theta}$, is the product of a power and an exponential function in $1/\sigma^2$. Recognizing a connection with the integral needed to normalize the kernel of an inverse gamma density suggests how to extend the results to the case of non-conjugate prior distributions. The next two corollaries make this extension, placing the focus on the tails of the prior distribution.

The corollaries assume independence between $\boldsymbol{\theta}$ and $\sigma^2$, and so we consider their tail behavior separately. Let $\pi_{11}$ denote a (proper) prior distribution on $\boldsymbol{\theta}$, and let $\mathcal{F}_1$ be the family of all nondegenerate multivariate normal distributions on $\mathcal{R}^k$. Corollary 3.2 distinguishes between priors that are thick-tailed with respect to $\mathcal{F}_1$ and those that are not. Let $\pi_{12}$ denote a (proper) prior distribution on $\sigma^2$, and let $\mathcal{F}_2$ be the family of all inverse gamma distributions, $IG(\alpha, \beta)$, $\alpha > 0$, $\beta > 0$. Exploiting the connection mentioned in the previous paragraph, the proof of Theorem 3.1 shows that conditions $(a), (a'), (c)$ and $(c')$ determine the integrability (or lack thereof) of a certain function of $\sigma^2$ in a neighborhood of zero. For $\sigma^2$ going to infinity, a suitable number of observations guarantees integrability. Thus, the corollaries focus on the tail for $\sigma^2$ near 0, or the tail for the precision, $1/\sigma^2$, tending to $\infty$. A distribution, $\pi_{12}$ which is thick-tailed with respect to $\mathcal{F}_2$ has the property that $\lim_{\sigma^2\to 0} \pi_{12}(\sigma^2)/\pi_{02}(\sigma^2) = \infty$, for all $\pi_{02} \in \mathcal{F}_2$; a distribution that is thin-tailed with respect to $\mathcal{F}_2$ satisfies $\lim_{\sigma^2\to 0} \pi_{12}(\sigma^2)/\pi_{02}(\sigma^2) = 0$, for all $\pi_{02} \in \mathcal{F}_2$.

Before proceeding, we summarize the notational conventions just introduced and the assumptions common to both corollaries.

1. $\mathcal{F}_1$ denotes the family of all nondegenerate multivariate normal distributions on $\mathcal{R}^k$.
2. $\mathcal{F}_2$ denotes the family of all inverse gamma distributions.
3. $\boldsymbol{\theta}$ and $\sigma^2$ are assumed to be independent.
4. $\pi_{11}$ denotes a prior distribution for $\boldsymbol{\theta}$ having full support $\mathcal{R}^k$.
5. $\pi_{12}$ denotes a prior distribution for $\sigma^2$ such that $\int (\sigma^2)^{(n-rI)/2}\pi_{12}(\sigma^2)\, d\sigma^2 < \infty$.
6. $\lambda_1 \le \cdots \le \lambda_I$ denote the eigenvalues of $H_{\mathcal{I}}$ assumed to satisfy $\lambda_i \ne 1/r$, for all $i = 1, \ldots, I$.

The first corollary deals with thick-tailed prior distributions $\pi_{12}$ on $\sigma^2$ and covers the case of all proper prior distributions $\pi_{11}$ on $\boldsymbol{\theta}$.

**Corollary 3.1.** *Assume 1.–6. above and let $\pi_{12}(\sigma^2)$ be thick-tailed with respect to $\mathcal{F}_2$. If $\lambda_I < 1/r$ and $\mathrm{RSS}^*_{\setminus\mathcal{I}}(r) > 0$, then the case-deleted weight function has finite $r$th moment with respect to the full posterior distribution. On the other hand, if $\lambda_I > 1/r$ or $\mathrm{RSS}^*_{\setminus\mathcal{I}}(r) < 0$, then the $r$th moment of the case-deleted weight function is infinite.*

The next corollary applies to thin-tailed distributions $\pi_{12}(\sigma^2)$. It provides only a sufficient condition if $\pi_{11}(\boldsymbol{\theta})$ is thin-tailed and necessary and sufficient conditions if $\pi_{11}(\boldsymbol{\theta})$ is thick-tailed with respect to $\mathcal{F}_1$.

**Corollary 3.2.** *Assume 1.–6. above and let $\pi_{12}(\sigma^2)$ be thin-tailed with respect to $\mathcal{F}_2$. If $\lambda_I < 1/r$, then the case-deleted weight function has finite $r$th moment with respect to the full posterior distribution. If $\pi_{11}(\boldsymbol{\theta})$ is thick-tailed with respect to $\mathcal{F}_1$, then $\lambda_I > 1/r$ implies that the case-deleted weight function has infinite $r$th moment.*

If $\lambda_I > 1/r$ and both the prior $\pi_{11}$ on $\boldsymbol{\theta}$ and the prior $\pi_{12}$ on $\sigma^2$ are thin-tailed, we cannot draw any conclusions about the finiteness of the full posterior $r$th moment of $w_{\setminus\mathcal{I}}(\boldsymbol{\theta}, \sigma^2)$ as shown in the following example.

**Example 3.1.** Consider the univariate regression model $y_j \sim N(\theta x_j, \sigma^2)$ with no intercept and with prior distribution on $\theta$, $\pi_{11}(\theta) \propto \exp\{-(\theta-\theta_0)^4\}$. Suppose we observe a sample with $i$th leverage $h_{ii} = 1/2 + 1/\sum_{j=1}^n x_j^2$. Although $h_{ii} > 1/2$, if the prior distribution on $\sigma^2$ is $\pi_{12}(\sigma^2) \propto \exp\{-(\sigma^2)^{-2} - \sigma^2\}$, the posterior second moment $E(w^2_{\setminus i}(\theta, \sigma^2)|\boldsymbol{y})$ is finite. On the other hand, if the prior distribution on $\sigma^2$ is $\pi_{22}(\sigma^2) \propto \exp\{-(\sigma^2)^{-3/2} - \sigma^2\}$, then $E(w^2_{\setminus i}(\theta, \sigma^2)|\boldsymbol{y}) = \infty$. Both prior distributions $\pi_{12}$ and $\pi_{22}$ are thin-tailed with respect to $\mathcal{F}_2$.

Finally, consider the case of a prior distribution $\pi_{11}(\boldsymbol{\theta})$ having bounded support. Arguing as in the proof of Corollary 3.2, it is easy to verify that the $r$th moment of $w_{\setminus\mathcal{I}}$, $E(w^r_{\setminus\mathcal{I}}(\boldsymbol{\theta},\sigma^2)|\boldsymbol{y})$, always exists if $\pi_{12}(\sigma^2)$ is thin-tailed with respect to $\mathcal{F}_2$. On the other hand, if $\pi_{12}(\sigma^2)$ is either in $\mathcal{F}_2$ or is thick-tailed with respect to $\mathcal{F}_2$, the finiteness of $E(w^r_{\setminus\mathcal{I}}(\boldsymbol{\theta},\sigma^2)|\boldsymbol{y})$ depends essentially on the value of $\text{RSS}^*_{\setminus\mathcal{I}}$.
More precisely, let $M = \min_{\boldsymbol{\theta}\in\,\text{support}(\pi_{11})} \boldsymbol{\theta}^T(X^TX - rX_{\mathcal{I}}X^T_{\mathcal{I}})\boldsymbol{\theta} - 2(\boldsymbol{y}^TX - r y^T_{\mathcal{I}}X^T_{\mathcal{I}})\,\boldsymbol{\theta}$. Then one can prove the following:

- if $\pi_{12}(\sigma^2) = IG(\alpha,\beta)$, then $\text{RSS}^*_{\setminus\mathcal{I}} > -(2/\beta + M)$ and $n/2+\alpha > rI/2$ imply $E(w^r_{\setminus\mathcal{I}}(\boldsymbol{\theta},\sigma^2)|\boldsymbol{y}) < \infty$, while $\text{RSS}^*_{\setminus\mathcal{I}} < -(2/\beta + M)$ or $n/2+\alpha \le rI/2$ implies $E(w^r_{\setminus\mathcal{I}}(\boldsymbol{\theta},\sigma^2)\,|\,\boldsymbol{y}) = \infty$;

- if $\pi_{12}(\sigma^2)$ is thick-tailed with respect to $\mathcal{F}_2$, then $\text{RSS}^*_{\setminus\mathcal{I}}(r) > -M$ and $\int(\sigma^2)^{-(n-rI)/2}\pi_{j2}(\sigma^2)\ d\sigma^2 < \infty$ imply $E(w^r_{\setminus\mathcal{I}}(\boldsymbol{\theta},\sigma^2)|\boldsymbol{y}) < \infty$, while $\text{RSS}^*_{\setminus\mathcal{I}}(r) < -M$ or $\int(\sigma^2)^{-(n-rI)/2}\pi_{j2}(\sigma^2)\ d\sigma^2 = \infty$ implies $E(w^r_{\setminus\mathcal{I}}(\boldsymbol{\theta},\sigma^2)|\boldsymbol{y}) = \infty$.

## 4. A nonlinear model

To illustrate some of the issues that arise when the fitted model is nonlinear, we revisit a Bayesian analysis of the Puromicyn data presented in [17]. The data come from a biochemical reaction and are described in [5], p. 425. For a group of cells not treated with the drug Puromycin, there are $n = 11$ measurements of the initial velocity of a reaction, $V_i$, obtained when the concentration of the substrate was set at a given positive value, $c_i$. The observations are recorded in Table 1 and plotted in Figure 1. The Bayesian model fit in [17] assumes a non linear regression of velocity on concentration given by the Michaelis-Menten (MM) relation:

$$\text{E}(V_i) = (mc_i)/(\kappa + c_i). \tag{4.1}$$

According to this relation, when the concentration of the substrate equals the Michaelis parameter, $\kappa$, the velocity reaches half of its maximal value, $m$, which is also the limiting velocity as the concentration goes to infinity.

Following [17], we model the $n$ observations as independent realizations from normal distributions with means given by Equation (4.1) and common variance $\sigma^2$. All three parameters $m$, $\kappa$, and $\sigma^2$ are constrained to be positive and their prior distribution is specified as $\pi(m,\kappa,\sigma^2) = \pi_1(m,\sigma^2)\,\pi_2(\kappa)$, with $\pi_1(m,\sigma^2) \propto 1/\sigma^2$ representing a noninformative prior density for $(m,\sigma^2)$ and $\pi_2$ representing a proper prior density for $\kappa$ such that $\int_0^\infty \kappa\,\pi_2(\kappa)\,d\kappa < \infty$. This requirement guarantees that the posterior is proper.

TABLE 1
*The Puromycin Data and Related Case-Deleted Quantities. The bottom row contains the moment index $r^*$, i.e., the least upper bound on the value of $r$ such that $E(w^r_{\setminus i}(m,\sigma^2,\kappa))|\boldsymbol{v}) < \infty$.*

| Case No. $i$ | 1 | 2 | 3 | 4 | 5 | 6 | 7 | 8 | 9 | 10 | 11 |
|---|---|---|---|---|---|---|---|---|---|---|---|
| Concentration | 0.02 | 0.02 | 0.06 | 0.06 | 0.11 | 0.11 | 0.22 | 0.22 | 0.56 | 0.56 | 1.1 |
| Velocity | 67 | 51 | 84 | 86 | 98 | 115 | 131 | 124 | 144 | 158 | 160 |
| $\sum_{j\notin\{i\}} c_j^2 - c_i^2$ | 1.97 | 1.97 | 1.96 | 1.96 | 1.94 | 1.94 | 1.87 | 1.87 | 1.34 | 1.34 | −0.45 |
| moment index $r^*$ | 1.59 | 2.79 | 4.48 | 5.17 | 2.86 | 5.19 | 6.38 | 5.26 | 3.77 | 2.81 | 1.32 |

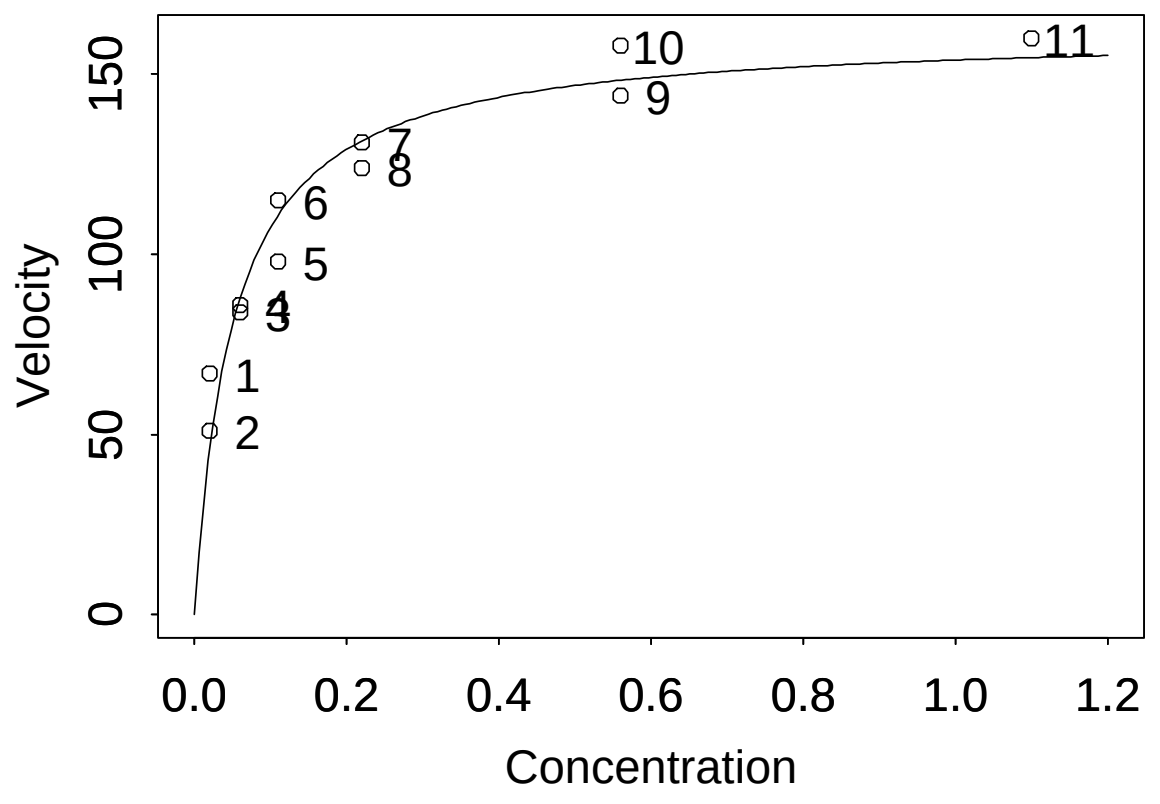

Fig 1. *Scatterplot of the Puromycin Data Set. The curve represents a fit of the expected velocity based on the posterior means of* $m$ *and* $\kappa$.

The MM model is, conditional on $\kappa$, a linear regression model with no intercept and covariate $x_i(\kappa) := c_i/(\kappa + c_i)$, for $i = 1,\dots,n$. Thus, for fixed $\kappa$, if the support of $m$ were $\mathcal{R}^1$, we could apply Theorem 3.1(ii). The case-deleted weight function is $w_{\setminus\mathcal{I}}(m,\kappa,\sigma^2) = (\sigma^2)^{I/2}\exp\{\sum_{i\in\mathcal{I}}[v_i - m x_i(\kappa)]^2/(2\sigma^2)\}$. Noting that $x_i(\kappa)$, $h_{ii}$ and $e_i$ are continuous functions of $\kappa$, we see that when Theorem 3.1(ii) indicates an infinite conditional $r$th moment at some value $\kappa_0$, it also indicates an infinite conditional moment in an open interval about $\kappa_0$. If the prior on $\kappa$ has full support, this interval receives positive posterior probability, and so the unconditional $r$th moment is infinite. The analog of Theorem 3.1 for the Michaelis-Menton model will impose conditions on leverage, sample size and residual.

The unconditional $r$th moment may be infinite for a different reason: the finite conditional $r$th moments may integrate to infinity. Thus, the conditions will need to be strengthened to ensure a finite $r$th moment. To avoid this route to infinity, the conditions on leverage and residual are applied uniformly in $\kappa$. Finally, an apparent infinite moment will sometimes be finite due to the restriction on the support of $m$.

Define the conditional design matrix $X(\kappa)$. Proceeding as in Section 3, define the matrix $H_{\mathcal{I}}(\kappa)$, and concentrate on its largest eigenvalue. The conditional leverage, $l(\mathcal{I},\kappa) = \sum_{i\in\mathcal{I}} x_i^2(\kappa)/\sum_{i=1}^n x_i^2(\kappa)$, is the only non-zero eigenvalue. The condition on the residual can be expressed in terms of simpler functions which will prove useful later. Define $A(\mathcal{I},r,\kappa) = \sum_{i\notin\mathcal{I}} x_i^2(\kappa) - (r-1)\sum_{i\in\mathcal{I}} x_i^2(\kappa)$, $B(\mathcal{I},r,\kappa) = \sum_{i\notin\mathcal{I}} x_i(k)v_i - (r-1)\sum_{i\in\mathcal{I}} x_i(\kappa)v_i$, and $C(\mathcal{I},r) = \sum_{i\notin\mathcal{I}} v_i^2 - (r-1)\sum_{i\in\mathcal{I}} v_i^2$. The adjusted, conditional residual sum of squares is $\mathrm{RSS}^*_{\setminus\mathcal{I}}(r,\kappa) = C(\mathcal{I},r) - B^2(\mathcal{I},r,\kappa)/A(\mathcal{I},r,\kappa)$. The set of zeroes of $A(\mathcal{I},r,\kappa)$, with $\mathcal{I}$ and $r$ held fixed, contains at most $2(n-1)$ points, a set of Lebesgue measure 0, and so we need not worry about the apparent division by 0. One last quantity is needed to handle the partial support of $m$. Define $g(\mathcal{I},\kappa) = \sum_{i\in\mathcal{I}} x_i(\kappa)v_i / \sum_{i=1}^{n} x_i(\kappa)v_i$, the product of covariate and response summed over the deleted cases divided by the same quantity summed over all cases.

The results on the finiteness of $E(w^r_{\mathcal{I}}(m,\sigma^2,\kappa)|\boldsymbol{v})$ are summarized in the following theorem.

**Theorem 4.1.** *Let $\pi_2$ be a proper prior distribution on $\kappa$ such that $\int_0^\infty \kappa\,\pi_2(\kappa)\,d\kappa < \infty$.*
*Suppose that*

(*a*) *there exists a measurable set $\mathcal{N}$ such that $\pi_2(\mathcal{N}) = 0$ and* $\sup_{\kappa\in\mathcal{N}^c} l(\mathcal{I},\kappa) < 1/r$

*and*

(*b*) $n > rI + 1$.

*If, in addition,*

(*c*) $\inf_{\kappa\in\mathcal{N}^c}\{\mathrm{RSS}^*_{\setminus\mathcal{I}}(r,\kappa)\} > 0$

*or*

(*d*) $C(\mathcal{I},r) > 0$ *and* $\inf_{\kappa\in\mathcal{N}^c} g(\mathcal{I},\kappa) > 1/r$

*holds, then the case-deleted weight function $w_{\setminus\mathcal{I}}(m,\kappa,\sigma^2)$ has finite $r$th moment with respect to the full posterior $p(m,\kappa,\sigma^2)$.*
*On the other hand, either of the conditions*

(*e*) $A(\mathcal{I},r,\kappa)m^2 - 2B(\mathcal{I},r,\kappa)m + C(\mathcal{I},r) < 0$ *for all $(m,\kappa)$ in some non-negligible set*

*or*

(*f*) $n \le rI + 1$

*is sufficient for the $r$th moment of the importance sampling weight function to be infinite.*

*Remark* 4.1. The sufficient conditions of Theorem 4.1 are essentially necessary for the $r$th posterior moment of the case-deleted weight function to be finite. If there exists a non-negligible set of values of $\kappa$ such that

($a'$) $l(\mathcal{I},\kappa) > 1/r$

or

($c'$) $\mathrm{RSS}^*_{\setminus\mathcal{I}}(r,\kappa)\} < 0$ and ($d'$) $C(\mathcal{I},r) < 0$ or $g(\mathcal{I},\kappa) < 1/r$,

then condition ($e$) is satisfied, and vice versa. This is because $l(\mathcal{I},\kappa) > 1/r$ if and only if $A(\mathcal{I},r,\kappa) < 0$, $B^2(\mathcal{I},r,\kappa) - C(\mathcal{I},r)A(\mathcal{I},r,\kappa)$ is the discriminant of the quadratic equation $A(\mathcal{I},r,\kappa)m^2 - 2B(\mathcal{I},r,\kappa)m + C(\mathcal{I},r) = 0$, and $g(\mathcal{I},\kappa) < 1/r$ if and only if $B(\mathcal{I},r,\kappa) > 0$.

*Remark* 4.2. For a general $r > 1$, upper bounds for the leverages $l(\mathcal{I},\kappa)$ and lower bounds for the functions $g(\mathcal{I},\kappa)$ and for the marginal residual sums of squares $\mathrm{RSS}^*_{\setminus\mathcal{I}}(r,\kappa)$ are hard to derive analytically, but numerical verification of the conditions of Theorem 4.1 is rather simple. In fact, $l(\mathcal{I},\kappa) \neq 1/r$ if and only if $A(\mathcal{I},r,\kappa) \neq 0$ and $g(\mathcal{I},\kappa) \neq 1/r$ if and only if $B(\mathcal{I},r,\kappa) \neq 0$. Moreover, for $\kappa > 0$, $A(\mathcal{I},r,\kappa)$, $B(\mathcal{I},r,\kappa)$ and $C(\mathcal{I},r)A(\mathcal{I},r,\kappa) - [B(\mathcal{I},r,\kappa)]^2$ are continuous functions that approach zero as $\kappa$ goes to infinity and that can only have a finite number of extrema. The latter can be found among the real positive roots of polynomials of degree $2n-3$, $n-2$ and $2n-3$ respectively.

*Remark* 4.3. The strategy applied to the MM model applies to an array of models that are, conditional on some set of parameters, linear. We impose the leverage and residual conditions uniformly across the parameters that render the model nonlinear. Important classes of models are linear regression models that allow for Box-Cox transformation of the response variable and/or Box-Tidwell transformation of the explanatory variables.

The authors of [17] specified a $t$ distribution on 3 degrees of freedom restricted to $[0,+\infty)$ as a prior $\pi_2$ for the parameter $\kappa$ and fit the model to the Puromycin data using the program BUGS (see [22]). (Due to some technical restrictions of BUGS, they had to use approximations for some of the prior specifications.) They considered deletion of single cases and computed the corresponding case-deleted weight functions. They reported detailed estimation results based on the deletion of case 1, an observation that produces highly variable realized weight functions, and illustrated how a transformation based approach (the Importance Link Function method) can effectively reduce the variability of the weight functions and lead to improved estimation.

We discuss the implications of the results developed in this section on the analysis presented in [17]. We consider, as was done in [17], deletion of single observations and focus on the case $r=2$, so that the sample size condition $(b)$ of Theorem 4.1 is satisfied for all cases. An examination of the leverage condition $(a)$ shows that observation 11 has large leverage (for $\kappa=2$, $l(\mathcal{I},\kappa)=0.5065>1/r=1/2$), and so by Remark 4.1, the posterior variance of the case-deleted weight function for case 11 is infinite. All remaining cases have leverages bounded away from 0 and above strictly by $1/2$, and so satisfy condition $(a)$.

Turning to the residual conditions $(c)$ and $(d)$, we find that all observations other than 1 and 11 satisfy condition $(c)$, thus ensuring finite variances for their case-deleted weight function. For observation 1, the adjusted residual sum of squares is negative for values of $\kappa$ near 0.08, violating condition $(c)$. Condition $(d)$ is also violated since $\sup_{\kappa>0} g(1,\kappa)=0.05501<1/2$. Consequently, Theorem 4.1 implies that the case-deleted weight function for observation 1 has infinite variance.

In addition to $r=2$, we can examine other moments of the case-deleted weight functions. Table 1 displays, for every case-deletion $i$, the moment index $r^*$, i.e., the least upper bound on the value of $r$ for which the $r$th moment exists (see [7] and [8]). If the influence of the $i$th observation on the posterior distribution $p(m,\sigma^2,\kappa)$ is assessed by the $\chi^2$ divergence measure between the case-deleted and full posteriors: $\chi^2=\int[p_{\setminus i}(m,\sigma^2,\kappa)/p(m,\sigma^2,\kappa)-1]^2 p(m,\sigma^2,\kappa)\; dm\; d\sigma^2\; d\kappa$, then, as suggested in [26] and [27], we can estimate $\chi^2$ by means of the Monte Carlo sum appearing in Table 2. As indicated in Section 6, this estimator is asymptotically normal if $E(w^4_{\setminus i}(s)|\boldsymbol{y})<\infty$. According to the values of $r$ displayed in Table 1, a central limit theorem holds only for the estimators of $\chi^2$ corresponding to observations $3,4,6,7$, and 8.

TABLE 2

*Influence Measures. This table presents a selection of influence measures and sufficient conditions for their estimators to follow a central limit theorem (see Section 6). $KL$ represents Kullback-Liebler divergence, $L1$ is integrated $L_1$ loss, $L2$ is integrated $L_2$ loss, $\Delta 1$ is change in first moment of a parameter $\theta$, $\Delta 2$ is change in second moment of a parameter $\theta$, $Hel$ is Hellinger distance, $ChSq$ is chi-square distance, $CPO$ is the Conditional Predictive Ordinate, and $Bdd$ is a bounded function. As a shorthand for the notation introduced in Section 2, a subscript $m$ means that a function is evaluated at $z_m$ (e.g., $w_m = w(z_m)$, etc.). The symbol $L_{\mathcal{I}}(s)$ represents the likelihood of the observations in $\mathcal{I}$ evaluated at the point $s$, $L_{\backslash\mathcal{I}}$ represents the likelihood of the observations not in set $\mathcal{I}$, and $\pi$ represents the prior density. The expression $2+\delta$ in the table means that it is sufficient, for some $\delta > 0$, that $2+\delta$ moments exist. $\hat{R} = \sum_{m=1}^{M} w_m/M$. $\hat{C}$ is an estimator of $C = \int q(s)\,ds$. There are many estimators of $C$, with some based on a different simulation than that used to fit the model. In lines 2 and 3 of the table, we assume that $\hat{C}$ is sufficiently well behaved that it does not prevent the estimators from following central limit theorems. In line 8 of the table, we presume that $w(s) = 1/L_{\mathcal{I}}(s)$.*

| Meas | Estimand | Estimator | Mom's | Adjmnt | Adj-Mom's |
|---|---|---|---|---|---|
| $KL$ | $\int \log(\frac{p(s)}{p_{\backslash\mathcal{I}}(s)})p_{\backslash\mathcal{I}}(s)\,ds$ | $-\hat{R}^{-1}\sum_{m=1}^{M} w_m \log(w_m)/M - \log(\hat{R})$ | $2+\delta$ | n.a. | n.a. |
| $L1$ | $\int \lvert p_{\backslash\mathcal{I}}(s) - p(s)\rvert p_{\backslash\mathcal{I}}(s)\,ds$ | $\hat{C}^{-1}\hat{R}^{-1}\sum_{m=1}^{M} q_m w_m \lvert\hat{R}^{-1}w_m - 1\rvert/M$ | 2 | $\pi^2 L_{\backslash\mathcal{I}}^2$ | 2 |
| $L2$ | $\int (p_{\backslash\mathcal{I}}(s) - p(s))^2 p_{\backslash\mathcal{I}}(s)\,ds$ | $\hat{C}^{-2}\hat{R}^{-1}\sum_{m=1}^{M} q_m^2(\hat{R}^{-1}w_m - 1)^2 w_m/M$ | 2 | $\pi^4 L_{\backslash\mathcal{I}}^4$ | 4 |
| $\Delta 1$ | $\int \theta p_{\backslash\mathcal{I}}(s)\,ds - \int \theta p(s)\,ds$ | $\sum_{m=1}^{M} \theta_m(\hat{R}^{-1}w_m - 1)/M$ | 2 | $\theta^2$ | 2 |
| $\Delta 2$ | $\int \theta^2 p_{\backslash\mathcal{I}}(s)\,ds - \int \theta^2 p(s)\,ds$ | $\sum_{m=1}^{M} \theta_m^2(\hat{R}^{-1}w_m - 1)/M$ | 2 | $\theta^4$ | 2 |
| $Hel$ | $\int (\sqrt{p(s)} - \sqrt{p_{\backslash\mathcal{I}}(s)})^2\,ds$ | $2 - 2\sqrt{\hat{R}^{-1}}\sum_{m=1}^{M}\sqrt{w_m}/M$ | 2 | n.a. | n.a. |
| $ChSq$ | $\int (\frac{p_{\backslash\mathcal{I}}(s)}{p(s)} - 1)^2 p(s)\,ds$ | $\sum_{m=1}^{M}(\hat{R}^{-1}w_m - 1)^2/M$ | 4 | n.a. | n.a. |
| $CPO$ | $\int L_{\mathcal{I}}(s) p_{\backslash\mathcal{I}}(s)\,ds$ | $\hat{R}^{-1}$ | 2 | n.a. | n.a. |
| $Bdd$ | $\int g(s) p_{\backslash\mathcal{I}}(s)\,ds$ | $\sum_{m=1}^{M}\hat{R}^{-1}g_m w_m/M$ | 2 | n.a. | n.a. |

## 5. Bayesian logistic regression

We now switch our focus to generalized linear models, concentrating on the study of a logistic regression model. Assume that, for each of $n$ subjects, we have available a $k \times 1$ vector of covariate information, $\boldsymbol{x}_i$, and we observe a 0-1 outcome, $Y_i$. Suppose that the $Y_i$ are independently distributed as Bernoulli random variables taking on value 1 with probability $p_i = \exp\{\boldsymbol{\beta}^T \boldsymbol{x}_i\}/[1 + \exp\{\boldsymbol{\beta}^T \boldsymbol{x}_i\}]$. The case-deleted weight function is proportional to

$$w_{\setminus \mathcal{I}}(\boldsymbol{\beta}) = \frac{q_{\setminus \mathcal{I}}(\boldsymbol{\beta})}{q(\boldsymbol{\beta})} = \prod_{i \in \mathcal{I}} \frac{1 + \exp\{\boldsymbol{\beta}^T \boldsymbol{x}_i\}}{\exp\{\boldsymbol{\beta}^T \boldsymbol{x}_i y_i\}}. \tag{5.1}$$

The following theorem covers prior distributions with the exponential tails that match the logistic regression likelihood. Subsequent corollaries cover thinner and thicker tailed prior distributions.

**Theorem 5.1.** *Let the data follow the logistic regression model just described, and assume that we have a prior distribution for $\boldsymbol{\beta}$ with density proportional to $\pi(\boldsymbol{\beta}) = \exp\{-\epsilon|\boldsymbol{\beta}^T|\mathbf{1}\}$, where $\epsilon > 0$ is given and $|\boldsymbol{\beta}^T|\mathbf{1} := \sum_{j=0}^{k-1} |\beta_j|$. Define*

$$h(\boldsymbol{\beta}, r, \epsilon) = \boldsymbol{\beta}^T \Bigg( \sum_{i \notin \mathcal{I}} \boldsymbol{x}_i y_i - (r-1) \sum_{i \in \mathcal{I}} \boldsymbol{x}_i y_i - \\ - \sum_{i \notin \mathcal{I}} \boldsymbol{x}_i I(\boldsymbol{\beta}^T \boldsymbol{x}_i > 0) + (r-1) \sum_{i \in \mathcal{I}} \boldsymbol{x}_i I(\boldsymbol{\beta}^T \boldsymbol{x}_i > 0) \Bigg) - \epsilon |\boldsymbol{\beta}^T| \mathbf{1}.$$

*If, for all vectors $\boldsymbol{\beta}$ such that $|\boldsymbol{\beta}^T|\mathbf{1} = 1$, $h(\boldsymbol{\beta}, r, \epsilon) < 0$, then the case-deleted weight function $w_{\setminus \mathcal{I}}(\boldsymbol{\beta})$ has finite $r$th moment with respect to the full posterior $p(\boldsymbol{\beta})$. If, for some vector $\boldsymbol{\beta}$ such that $|\boldsymbol{\beta}^T|\mathbf{1} = 1$, $h(\boldsymbol{\beta}, r, \epsilon) > 0$, then the case-deleted weight function has infinite $r$th moment.*

The theorem can be applied to prior distributions proportional to $\exp\{-|\boldsymbol{\beta}^T|\boldsymbol{\epsilon}\}$, where $\boldsymbol{\epsilon}$ is a vector of positive numbers. In this instance, a rescaling of the covariates to obtain a prior with a single real $\epsilon$ results in the type of prior for which the theorem is stated.

The theorem may be strengthened somewhat by explicitly considering the case of $\max_{\boldsymbol{\beta}:|\boldsymbol{\beta}^T|\mathbf{1}=1} h(\boldsymbol{\beta}, r, \epsilon) = 0$, although the statement of precise conditions under which the case-deleted $r$th moment is infinite becomes messy. The conditions in Theorem 5.1 are easy to check since the maximum of $h(\boldsymbol{\beta}, r, \epsilon)$ may be found via linear programming methods.

As in the case of the linear model, we will investigate the $r$th moment of the case-deleted weight function under thick-tailed and thin-tailed prior distributions. The main tool for the proofs is, once again, Lemma 2.1. The first corollary deals with thick-tailed distributions.

**Corollary 5.1.** *Let the prior distribution on $\boldsymbol{\beta}$ have thick tails with respect to the class of distributions $\mathcal{F} = \{\pi(\boldsymbol{\beta}) : \pi(\boldsymbol{\beta}) = c\exp(-\epsilon|\boldsymbol{\beta}^T|\mathbf{1})$ and $\epsilon > 0\}$. Then,* *if* $h(\boldsymbol{\beta}, r, 0) < 0$ *for all* $\boldsymbol{\beta}$ *such that* $|\boldsymbol{\beta}^T|\mathbf{1} = 1$*, the case-deleted weight function has finite* $r$*th moment* $(r > 0)$ *with respect to the full posterior* $p(\boldsymbol{\beta})$*. If, for some vector* $\boldsymbol{\beta}$ *such that* $|\boldsymbol{\beta}^T|\mathbf{1} = 1$*,* $h(\boldsymbol{\beta}, r, 0) > 0$*, then the case-deleted weight function has infinite* $r$*th moment.*

The next corollary applies to thin-tailed distributions.

**Corollary 5.2.** *Let the prior distribution on* $\boldsymbol{\beta}$ *have thin tails with respect to the class of distributions* $\mathcal{F} = \{\pi(\boldsymbol{\beta}) : \pi(\boldsymbol{\beta}) = c\exp(-\epsilon|\boldsymbol{\beta}^T|\mathbf{1})$ *and* $\epsilon > 0\}$*. Then the case-deleted weight function has finite* $r$*th moment with respect to the full posterior* $p(\boldsymbol{\beta})$ *for all* $r > 0$*.*

### 5.1. Applying the corollaries

The preceding corollaries enable us to determine quickly whether the case-deleted weight function has finite or infinite $r$th moment. Consider an arbitrary logistic regression model where the prior distribution on $\boldsymbol{\beta}$ is taken to be the normal distribution with mean $\boldsymbol{\beta}_0$ and variance $\Sigma$, with $\Sigma$ of full rank. This distribution is thin-tailed with respect to the family of prior distributions used in Theorem 5.1. To verify this, write the ratio of priors, with $g$ representing the normal prior density and $f$ representing the prior density under a member of the exponential-tailed class:

$$
\begin{aligned}
\frac{g(\boldsymbol{\beta})}{f(\boldsymbol{\beta})} &= \frac{(2\pi)^{-k/2}|\Sigma|^{-1/2}\exp(-(\boldsymbol{\beta}-\boldsymbol{\beta}_0)^T\Sigma^{-1}(\boldsymbol{\beta}-\boldsymbol{\beta}_0)/2)}{c\exp(-\epsilon|\boldsymbol{\beta}^T|\mathbf{1})} \\
&\leq (2\pi)^{-k/2}|\Sigma|^{-1/2}c^{-1}\exp(-1/(2\lambda_1)(\boldsymbol{\beta}-\boldsymbol{\beta}_0)^T(\boldsymbol{\beta}-\boldsymbol{\beta}_0) + \epsilon|\boldsymbol{\beta}^T|\mathbf{1}) \\
&= (2\pi)^{-k/2}|\Sigma|^{-1/2}c^{-1}\exp(-1/(2\lambda_1)\sum_{i=1}^{k}(\beta_i-\beta_{0i})^2 + \sum_{i=1}^{k}|\beta_i|\epsilon),
\end{aligned}
$$

where $\lambda_1$ is the largest eigenvalue of $\Sigma$. Applying Corollary 5.2 with a normal prior distribution, we find that all positive moments of the case-deleted weight function are finite. This result holds, even if all of the cases are deleted.

Suppose instead that the prior distribution on $\boldsymbol{\beta}$ is taken to be a multivariate $t$ distribution with $\nu$ degrees of freedom, location vector $\boldsymbol{\beta}_0$ and scale matrix $\Sigma$, with $\Sigma$ of full rank. This $t$ distribution is thick-tailed with respect to the family of prior distributions used in Theorem 5.1. A formal verification of this follows from an examination of the ratio of prior density functions. To establish finiteness or infiniteness of the case-deleted moments, use Theorem 5.1 with $\epsilon = 0$.

We note that Theorem 5.1 can be of help in establishing whether or not the $r$th moment of the case-deleted weight function will be infinite, even when the prior distribution is improper. If the prior density for $\boldsymbol{\beta}$ is uniform on $\mathcal{R}^k$, for example, we merely apply the theorem with $\epsilon = 0$. The conclusion of a finite case-deleted $r$th moment is conditional upon the propriety of the posterior distribution. This propriety is not guaranteed, as use of the uniform prior distribution may lead to an improper posterior distribution (see [12] and [18]). However, since the weight function $w_{\setminus\mathcal{I}}(\boldsymbol{\beta})$ in Equation (5.1) always exceeds one, if the first moment of the case-deleted weight function is finite, so is the normalizing constant for the posterior: the posterior distribution is proper if any case-deleted weight function has finite first moment.

## 6. Central limit theorems

The previous sections provide results that enable us to calculate the number of moments which exist for the case-deleted weight function. The results apply to classes of prior distributions, and so can be quickly used to establish asymptotic normality of the importance sampling estimator $\hat{\mathrm{E}}_{p_{\setminus\mathcal{I}}}[g(s)]$ given in Equation (2.1) and of the related estimator $\hat{\mathrm{E}}^*_{p_{\setminus\mathcal{I}}}[g(s)]$. In this section, we indicate how these results apply to a variety of measures of case influence. We also present two techniques which are generally useful for applying the results.

Central limit theorems (CLTs) for importance sampling estimators when the parameter vectors $s$ are generated as i.i.d. samples or arise from a uniformly ergodic Markov chain, are described in [13] and [24], respectively. Under either source for the sample, the estimator $\hat{\mathrm{E}}^*_{p_{\setminus\mathcal{I}}}[g(s)]$ is asymptotically normal if and only if

$$\int w^2_{\setminus\mathcal{I}}(s)g^2(s)p(s|\boldsymbol{y})\;ds < \infty. \tag{6.1}$$

Sufficient conditions for $\hat{\mathrm{E}}_{p_{\setminus\mathcal{I}}}[g(s)]$ to be asymptotically normal are that condition (6.1) holds and that

$$\int w^2_{\setminus\mathcal{I}}(s)p(s|\boldsymbol{y})\;ds < \infty. \tag{6.2}$$

These conditions are explicitly presented for i.i.d. samples in [13]. A slight technical extension of the CLT in [24] helps to establish the result for ergodic samples. The extension consists of an application of the Cramer-Wold device to establish the joint asymptotic normality of the estimator of the normalizing constant for the weight function and of an estimator proportional to $\hat{\mathrm{E}}^*_{p_{\setminus\mathcal{I}}}[g(s)]$, followed by an application of the delta method (e.g., see [9]).

The first technique for establishing a CLT recognizes that the $g^2(s)$ term in the integral in condition (6.1) can be grouped with $p(s|\boldsymbol{y})$, yielding, say, $p^*(s|\boldsymbol{y})$. The quantity $p^*(s|\boldsymbol{y})$ is the formal posterior distribution for $s$ given the data, provided that it is integrable. It corresponds to a proper Bayesian analysis with $g^2(s)$-adjusted prior distribution proportional to $g^2(s)\pi(s)$, provided that $0 < \int g^2(s)\pi(s)\;ds < \infty$. We note that this integral is against the prior distribution, and so is typically easy to evaluate.

To facilitate application of the theorems we wish to preserve full support of the function-adjusted prior distribution. To check the asymptotic normality of $\hat{\mathrm{E}}^*_{p_{\setminus\mathcal{I}}}[g(s)]$ we need only verify condition (6.1), provided $g(s)$ is never equal to zero so that the $g^2(s)$-adjusted prior has full support. In all other cases we act as if the prior distribution had density $(1+g^2(s))\pi(s)$. This preserves full support of the function-adjusted prior distribution in case $g(s)$ is not always different from zero. This also allows us to verify at once conditions (6.1) and (6.2) when we wish to determine if $\hat{\mathrm{E}}_{p_{\setminus \mathcal{I}}}[g(s)]$ is asymptotically normal.

The second technique that we find useful is to establish the finiteness of integrals in case-deleted posteriors for a bounding function which then implies finiteness for interesting classes of functions. The relation $\log^2(x) \leq (C_\epsilon + x^{-\epsilon} + x^\epsilon)^2$ for some constant $C_\epsilon$ and all $x > 0$ connects moments of the case-deletion weight function to finiteness of integrals for several influence measures. The moment generating function is also a useful bounding function. Hence, we consider $g(s) = \exp(s^T \boldsymbol{t})$ for all $\boldsymbol{t}$ in some open neighborhood of $\mathbf{0}$, say, $U$. If $\int w^2_{\setminus \mathcal{I}}(s)\exp(s^T\boldsymbol{t})p(s|\boldsymbol{y})\,ds < \infty$, for all $\boldsymbol{t} \in U$, then, condition (6.1) is satisfied for any polynomial in $s_1^{n_1}\cdots s_k^{n_k}$ and any constant. We note that condition (6.1) implies condition (6.2) and the CLT applies to the importance sampling estimators of any mixed and marginal moment of $s$.

Formal Bayesian techniques that describe the influence of a set of cases on an analysis focus on a one-dimensional summary of the difference between the case-deleted posterior distribution and the full posterior distribution. Bayesian measures of model fit focus on case-deleted measures of predictive accuracy and cross-validation. A plethora of summaries exist. In this subsection, we show how our results can be used to verify that a CLT holds for the summaries estimated on the basis of a Monte Carlo sample. We illustrate this point with a discussion at the end of Example 6.1 concerning estimation of the conditional predictive ordinate (CPO).

This approach can be applied to many Bayesian case influence measures. Table 2 contains a summary of results. Each row of the table corresponds to a measure of influence. The measure is given under the column headed Estimand, and a formula for estimation is given under the heading Estimator. The last three columns present sufficient conditions for the estimator to follow a CLT. The column headed Mom's gives a number of moments of the case-deleted weight function; the column headed Adjmnt presents the function used to adjust the prior distribution, if needed, and the column headed Adj-Mom's gives a number of moments of the case-deleted weight function against the function-adjusted prior distribution. If the given numbers of moments and adjusted moments both exist, then a CLT holds for the estimator.

### *6.1. Examples*

**Example 6.1.** This example illustrates the practical use of the results presented in Section 3. We fit a linear model to data assembled by the authors of [20] to investigate growth rates across mammalian species. Gestational time is known to be an important factor in determining growth rate. The data set contains 96 entries with complete information on growth rates and possibly related covariates for mammalian species. There is one marsupial that we excluded from our analysis. Three of the remaining species exhibit delayed implantation, a phenomenon by which the blastocyst, after reaching the uterus, remains dormant and unattached to the uterine lining for an extended period of time. An examination of the covariate gestation time (in days) led us to conclude that the recorded gestational time for the grizzly and polar bears–ursus arctos and thalarctos maritimus–included the preimplantation time while the recorded gestational time for the nine-banded armadillo–dasypus novemcinctus–did not. This last gestational time was adjusted to include preimplantation. After this adjustment, the recorded gestation time for each species included in the analysis covers the time from egg fertilization to birth.

The response variable is a species' advancement, defined as the ratio of neonatal to adult body weight. We built a linear model including an intercept and three covariates: the natural logarithms of gestation time, litter size, and adult body weight (centering all three covariates around their respective means). The least squares fit of this model yields a multiple $R$-square of 0.4344. Based on a Bayesian analysis with noninformative prior distributions for the model parameters, the 95% highest posterior density (hpd) intervals for the coefficients for log litter size and log body weight include only negative values while the 95% hpd interval for the coefficient for log gestation time includes only positive values. This indicates that heavier species, species with larger litter sizes, and species with shorter gestation times give birth to relatively immature offspring.

We use the theoretical results of the preceding sections for three purposes: we examine the influence of a preselected group of cases on inference, we screen all groups of cases of a certain size for their influence, and we verify the stability of cross-validatory estimators of summary measures. First, consider the three species with delayed implantation. We interpret the moment index $r^*_{\mathcal{I}}$, i.e., the cut-off value for the existence or non-existence of the $r$th moment of the case deletion weights (see, [7] and [8]), as a measure of influence of the cases being excluded. This cut-off value is given by the minimum of the cut-off value $r^*_{a,\mathcal{I}}$ between the leverage conditions $(a)$ and $(a')$ and the cut-off value $r^*_{c,\mathcal{I}}$ between the distance conditions $(c)$ and $(c')$. Dropping the three species leads to the values $r^*_{a,\mathcal{I}} = 4.74$ and $r^*_{c,\mathcal{I}} = 2.93$. Thus $r^*_{\mathcal{I}} = 2.93$ for this set of species. This number is small, suggesting that this group of cases is influential. A glance at Table 2 shows us that a central limit theorem will not hold for the chi-square distance, but that it will hold for the other measures listed in the table.

As with traditional measures of influence, we consider where our set of observations falls on the measures $r^*_{a,\mathcal{I}}$ and $r^*_{c,\mathcal{I}}$ with respect to other sets of similar size. We scanned all triples of species, computing cut-offs for each triple. Ordering the triples of dropped species according to their increasing values of $r^*_{a,\mathcal{I}}$, we found that the nine-banded armadillo belongs to 99 of the top 100 triples (all but the 31st), while the grizzly bear belongs to 2 of the top 100 triples (the 18th and the 99th), and the polar bear belongs to one of the top 100 triples (the 38th). Our three species in combination rank 1343rd out of the 138415 triples, with an $r^*_{a,\mathcal{I}}$ value of 4.74. Ordering the triples of dropped species according to their increasing values of $r^*_{c,\mathcal{I}}$, we find that both bear species belong to each of the top 93 triples and that the grizzly bear belongs to each of the top 167 triples. Dropping all three species with delayed implantation at once yields the 6th smallest value for $r^*_{c,\mathcal{I}}$. From this comparative analysis, we conclude that the three species with delayed implantation may be influential, the nine-banded armadillo due mainly to its leverage and the two bear species due mainly to their outlyingness. This set of three species stands out, as there is a common underlying factor that may differentiate them from the other species.

Pursuing the potential influence of our triple of cases, we examine whether inclusion of the dormant period in the total gestation time affects the conclusions that we draw from the model. To answer this question we adjusted the gestation times of these species to account only for the period of actual development and reconsidered the linear model described above. The least squares fit now yields a multiple $R$-square of 0.5267. The 95% hpd interval for the coefficient for log litter size now contains 0, suggesting possible simplification of the model, although the qualitative interpretations of the impact of species weight, litter size, and gestation time remain the same.

Repeating the earlier exercise of dropping triples of cases, we find that the leverage of the nine-banded armadillo is a little diminished, as it now enters only in 7 of the top 100 triples for $r^*_{a,\mathcal{I}}$. The grizzly bear has a little more leverage, as it now belongs to 4 of the top 100 triples (the 7th, 10th, 17th and 65th), while the polar bear belongs to just one of the top 100 triples (the 98th). The three species in combination rank 1916th with an $r^*_{a,\mathcal{I}}$ value of 5.24. The smallest value of $r^*_{c,\mathcal{I}}$, which now equals 3.65, is attained when the three species papio papio, ursus arctos, and thalarctos maritimus are dropped. Both bear species belong to each of the top 22 triples, the grizzly bear belongs to 96 of the top 100 triples and the polar bear belongs to 97 of the top 100 triples. Dropping all three species with delayed implantation at once yields the 23rd smallest value for $r^*_{c,\mathcal{I}}$.

Thus, the three species with delayed implantation are still influential when the model is fit to the adjusted gestation times, although the extent of their influence is slightly diminished. According to both analyses, the two bear species are highly influential, due mainly to their large residuals. This not only confirms the well known fact that bears have an unusually small advancement but also reveals that the dormant gestation period by itself cannot account for it. Quoting from a January 27, 2004 New York Times article (see [1]):

> Polar bears share with all bears an extreme disparity between the size of their mother, in the quarter-ton range, and that of a newborn cub—about a pound. “It’s dramatic trait in the bear family,” Dr. Peatkau said. “They are off the chart among placental mammals, and closer to marsupials like the kangaroo.”

Model fit is commonly assessed via $k$-fold cross validation. The data are partitioned at random into $k$ subsets of approximately equal sizes and each of the $k$ subsets is used in turn as a test set with the union of the remaining $k-1$ subsets serving as a training set. The model is fit to the data in the training sets and its predictions are compared to the actual values of the observations in the test sets by computing some measure of predictive ability averaged over the $k$ sets of predictions. In a Bayesian analysis, $CPO$ provides a measure of overall predictive ability. The cross-validated $CPO$ can be estimated with draws from the full posterior and importance sampling weights. However, as noted in Table 2, for a given partition, the central limit theorem will not hold if $r^*_{\mathcal{I}}$ drops below 2 when any of the $k$ subsets of observations is excluded.

To investigate how often the central limit theorem breaks down for $CPO$, we considered the case of 5-fold cross validation for the model fit to the data used in the first analysis and simulated 10,000 random partitions of the data into 5 subsets of size 19 each. For each split we computed five values of $r^*_{\mathcal{I}}$. Out of the total 10,000 simulated partitions, there were 658 partitions where $r^*_{\mathcal{I}}$ dropped below 2 for exactly one of the five case deletions and there was one partition where it dropped below 2 for two of the the five case deletions. The value of $r^*_{\mathcal{I}}$ never dropped below 2 for more than two of the five case deletions.

If it is established that, for a particular partition, no central limit theorem holds for importance sampling estimation of $CPO$, then the analyst must turn to other methods of estimation. For example, sampling from a mixture distribution with components given by the full posterior and by the case deleted posteriors conditional on those subsets for which $r^*_{\mathcal{I}} \leq 2$ ensures the existence of a central limit theorem for the estimate of $CPO$.

**Example 6.2.** The authors of [11] in their influential paper on Bayesian model selection/model averaging put a prior distribution over a collection of Bayesian linear models. There have been a host of extensions of their model, most of which are amenable to the treatment below. Formally, we describe a prior distribution having the form of Equation (3.2). The likelihood for the model follows Equation (3.1). The prior distribution on the error variance is $\sigma^2 \sim IG(\alpha, \beta)$. The prior distribution on the regression coefficients is described in two stages. At the first stage, there is an indicator vector of whether a regressor, $\theta_j$, "appears in" the model. The indicators are independent Bernoulli($p_j$) variates. If the regressor does not, then the conditional prior distribution on $\theta_j$ is $N(0, \tau^2)$ with small $\tau$; if the regressor does, then the conditional prior distribution on $\theta_j$ is $N(0, c\tau^2)$, with large $c > 1$. Marginalizing $p_j$, the prior distribution on an individual regressor is $\theta_j \sim (1 - p_j)N(0, \tau^2) + p_j N(0, c\tau^2)$. The resulting prior distribution remains absolutely continuous with respect to Lebesgue measure while effectively allowing regressors to be included in or excluded from the model.

The regression analysis is used to estimate the regression coefficients and the associated posterior expected loss. Pursuing a decision theoretic approach, we ask when the case-deleted importance sampling estimators follow CLTs. We use the standard sum-of-squared error loss, so that $L(\boldsymbol{\theta}, \boldsymbol{a}) = \sum_{j=0}^{k-1} (\theta_j - a_j)^2$. The Bayes action, $\boldsymbol{a}$, is the posterior mean vector. Here, we focus on the posterior expected loss. The posterior expected loss is $E[L(\boldsymbol{\theta}, \boldsymbol{a})|\boldsymbol{y}] = \sum_{j=0}^{k-1} Var(\theta_j|\boldsymbol{y})$, and then, for the asymptotic normality of $\hat{\mathrm{E}}_{p_{\setminus \mathcal{I}}}[g(s)]$ or $\hat{\mathrm{E}}^*_{p_{\setminus \mathcal{I}}}[g(s)]$, the function $g(\boldsymbol{\theta})$ to be considered is $g(\boldsymbol{\theta}) = \sum_{j=0}^{k-1} \theta_j^2$.

We now proceed with the technique. First, we verify that the function-adjusted prior distribution is proper. Since the prior distribution on the regression coefficients is a finite mixture of normals,

$$\int (1 + g^2(\boldsymbol{\theta}))\pi(\boldsymbol{\theta})\, d\boldsymbol{\theta} < \infty. \tag{6.3}$$

Next, we consider Theorem 3.1 as applied to the function-adjusted prior distribution with $r = 2$. If conditions $(a)$, $(b)$ and $(c)$ of the theorem hold for a particular case-deletion, then the case-deleted weight function have finite second moment, or equivalently, $\int w^2_{\setminus\mathcal{I}}(\boldsymbol{\theta}, \sigma^2)(1 + g^2(\boldsymbol{\theta}))p(\boldsymbol{\theta}, \sigma^2|\boldsymbol{y})\ d\boldsymbol{\theta}\ d\sigma^2 < \infty$, establishing conditions (6.1) and (6.2) and hence a CLT for the estimators $\hat{\mathrm{E}}_{p\setminus\mathcal{I}}[g(s)]$ and $\hat{\mathrm{E}}^*_{p\setminus\mathcal{I}}[g(s)]$. (Finiteness of the previous integral implies finiteness of $\int \sum_{j=0}^{k-1} w^2_{\setminus\mathcal{I}}(\boldsymbol{\theta}, \sigma^2)\theta_j^4 p(\boldsymbol{\theta}, \sigma^2|\boldsymbol{y})\ d\theta\ d\sigma^2$.) We note that conditions $(a), (b)$ and $(c)$ involve leverage, residuals from a least squares regression including all regressors and number of cases deleted. They do not directly involve the prior distribution, beyond the parameters $\alpha$ and $\beta$ of the prior on $\sigma^2$.

The impact of the prior distribution's tail behavior on decision rules is discussed in [2]. Robustness considerations suggest that it is often wise to use a prior distribution with thicker tails than the likelihood. For MCMC algorithms, a convenient replacement of the normal distribution is a $t$-distribution, see for example [4]. The technique used above can be directly applied and yields the same results when the prior distribution for $\theta_j$ is $(1-p_j)N(0, \tau^2)+p_jT(d, 0, c\tau^2)$, with the latter term in the mixture a $t$-distribution with $d > 4$ degrees of freedom, center 0 and scale $\tau^2$. The requirement $d > 4$ guarantees that condition (6.3) holds.

**Example 6.3.** The results of a study used to estimate the survival distribution for leukemia patients are presented in [10]. The response variable is survival time (from diagnosis), and explanatory variables are white blood cell count at diagnosis (WBC) and whether "Auer rods and/or significant granulature of the leukemic cells in the bone marrow at diagnosis" were present (AG positive) or absent (AG negative). The authors of [10] develop estimates of the survival distribution based upon presumed exponential distributions which are allowed to depend on the covariates. The authors of [6] dichotomize the survival times by defining a new response which indicates survival past 50 weeks. They analyze the data with the frequentist counterpart to the logistic regression model described in Section 5, where there are $k = 3$ covariates: an intercept, WBC and AG. The authors of [6] identify one case, a patient with a high WBC count and a survival time of more than 50 weeks, as having extremely large influence. They also note that altering the model (to predict survival based on log(WBC) and AG) can reduce the influence of the case.

We examine influence under a product of double-exponential prior distributions for $\boldsymbol{\beta}$. The distribution has scale parameter 10 in each direction (and hence a prior distribution with mean for $\boldsymbol{\beta}_i|(\boldsymbol{\beta}_i > 0)$ of 10). Case 15, diagnosed in [6] as an influential observation, is easily found to have an infinite variance for its case-deleted weight function. The value of the criterion $h(\boldsymbol{\beta}, 2, 0)$ is found to be $h((0, -1, 0)^T, 2, -0.1) = 45.15$. This value is well in excess of 0, and indicates that the choice of $\epsilon = 0.1$ for the prior distribution has little to do with why the case-deleted weight function has infinite variance. On the other hand, the value of the criterion in the positive direction for $\beta_2$ is less than 0, indicating that this tail of the distribution of $\beta_2$ is well-behaved. No other case results in an infinite variance for its case-deleted weight function.

For case 15 condition (6.2) does not hold and the estimators $\hat{\mathrm{E}}_{p\setminus\mathcal{I}}[g(s)]$ and $\hat{\mathrm{E}}^*_{p\setminus\mathcal{I}}[g(s)]$ are not asymptotically normal. Condition (6.2) holds for all remaining observations. Thus we can establish the asymptotic normality of their associated estimators by showing that condition (6.1) holds as well. We do this by using the bounding strategy described above showing that $h(\boldsymbol{\beta}, 2, \epsilon) < 0$, for all $\boldsymbol{\beta} : |\boldsymbol{\beta}^T|\mathbf{1} = 1$, implies the existence of an open neighborhood $U$ of $\mathbf{0}$ such that $\int w^2_{\setminus\mathcal{I}}(s)\exp(s^T\boldsymbol{t})p(s|\boldsymbol{y})\ ds < \infty$, for all $\boldsymbol{t} \in U$. Hence, it follows that $\int \exp\{h(\boldsymbol{\beta}, 2, \epsilon) + 2\boldsymbol{\beta}^T\boldsymbol{t}\}\ d\boldsymbol{\beta}$ is finite for all $\boldsymbol{t}$ in $U$, which in turn, arguing as in the proof of Theorem 5.1, implies that $\int w^2_{\setminus\mathcal{I}}(s)\exp(s^T\boldsymbol{t})p(s|\boldsymbol{y})\ ds < \infty$, for all $\boldsymbol{t} \in U$.

It is interesting to note that the analysis above is not strictly connected to the particular choice of the prior distribution as a product of double exponentials. Indeed, in light of Corollary 5.1, if the (proper) prior distribution on $\boldsymbol{\beta}$ is thick-tailed with respect to the family of product of double exponential distributions, then $h(\boldsymbol{\beta}, 2, 0) < 0$ for all $\boldsymbol{\beta} : |\boldsymbol{\beta}^T|\mathbf{1} = 1$ still implies both conditions (6.1) and (6.2). This is true, even when the noninformative prior distribution $\pi(\boldsymbol{\beta}) \sim 1$ is assigned. Finally, if $\pi(\boldsymbol{\beta})$ is thinner-tailed than any product of double exponentials, then conditions (6.1) and (6.2) are always satisfied.

## 7. Conclusions

The development of effective computational tools for fitting hierarchical models has spurred the growth of Bayesian data analysis. As with its classical counterpart, a complete Bayesian data analysis investigates sensitivity of inferences to changes in the data set, with particular consideration given to excluding observations from the analysis. This exclusion is most often accomplished through the use of importance sampling based on case-deleted weight functions. The theoretical results in Sections 3 through 5 provide conditions under which importance sampling estimators of various functionals will follow central limit theorems. Further results along these lines may be obtained for other likelihoods (particularly those in the exponential family) and for other specific model structures (as in Section 4). The techniques in Section 6 provide a simple means of verifying the conditions of the earlier theorems. We have found that the combination of these techniques and the theorems allow us to easily verify (or disprove) asymptotic normality of many estimators.

The results can be used to evaluate computational strategies. In many situations, computations can be hastened by sampling from a formal model that uses a nicely structured prior distribution–say $\pi_s(s)$–in place of the actual prior distribution, $\pi(s)$. This change may be motivated by the speed of programming conjugate calculations or by the speed of execution of the algorithm (e.g., see [17]) used to fit the model. With the altered model, inference is made through use of importance sampling with weights $w_p(s) = \pi(s)/\pi_s(s)$. When concerned about the effects of groups of cases, these importance sampling weights can be combined with the case-deletion weights to produce inference under the case-deleted posterior distribution. The weights are $w(s) = w_p(s)w_{\setminus\mathcal{I}}(s)$. Suppose that the weights due to the prior distribution have $r_p$ moments and the case-deletion weights have $r_{\mathcal{I}}$ moments (under the model with prior distribution $\pi_s$). Then a straightforward calculation shows that the combined weights have at least $(r_p^{-1} + r_{\mathcal{I}}^{-1})^{-1}$ moments. Thus, the suitability for quick and efficient data analysis based on the computational strategy where $\pi$ is replaced by $\pi_s$ for the sampling algorithm can be evaluated.

There is a strong connection between the tail of the prior distribution relative to the likelihood and the robustness of inference based on the model. Sentiment generally favors prior distributions with thicker tails than the likelihood. With a thick-tailed prior distribution, when there is a clash between likelihood and prior, inference is dominated by the likelihood (e.g., see [2], Chapter 4). Our preference is to select a prior distribution that reflects the analyst's beliefs. Often, this will be a thick-tailed prior distribution, leading to simplified conditions such as those in Corollaries 3.1 or 5.1. While our preference is to select the prior distribution on the basis of modeling considerations, we do note that the results of this paper could be used to select a prior with tails thin enough to guarantee existence of some targeted $r$ moments.

The results we derive apply to broad classes of models. As an example, the specification of the normal theory linear model in (3.1) and (3.2) can mask a much richer hierarchical model. The richer model may include further parameters–say $\gamma$–where the prior distribution on $\boldsymbol{\theta}$ depends on $\gamma$. As long as the likelihood is a function only of $\boldsymbol{\theta}$ and $\sigma^2$, the case-deleted weight function will also be a function of these parameters. The theorems are applied with the marginal prior distribution of $\boldsymbol{\theta}$ and $\sigma^2$. The prior specifications in [11] and [19] may be viewed in this light.

Models which combine different studies provide a less evident match for these theorems. A typical linear model used for such combination will allow the regression coefficients to vary from study to study. Such variation is captured with a hierarchical model that links the coefficients across studies by means of hyperparameters. The overall model can be expressed in graphical form as a hierarchical model. The advantage of the general conditions in the theorems that describe only the tail behavior of the prior distribution becomes apparent in this setting. For case deletions involving only one study, and referring to the notation of the previous paragraph, $\gamma$ includes the parameters specific to the other studies, the data specific to the other studies, and the hyperparameters. Thus the marginal prior distribution on $\boldsymbol{\theta}$ and $\sigma^2$ to be used in the theorems is the marginal distribution on these parameters, posterior to the data from the other studies. While this distribution is usually inaccessible in closed form, one can often verify that its tails behave like some (unspecified) normal distribution or that they are thicker than the class of normal distributions. This is sufficient for application of the theoretical results.

# APPENDIX

## Proof of Lemma 2.1

To prove the first part of the lemma, note that

$$S_1 = c_1^{-1} \int f(x)\pi_1(x)h(x)\,dx > c_1^{-1} \int f(x)B^{-1}\pi_0(x)h(x)\,dx$$
$$= c_1^{-1}B^{-1}c_0S_0 = \infty.$$

To prove the second part of the lemma, note that

$$S_1 = c_1^{-1} \int f(x)\pi_1(x)h(x)\,dx < c_1^{-1} \int f(x)b^{-1}\pi_0(x)h(x)\,dx$$
$$= c_1^{-1}b^{-1}c_0S_0 < \infty.$$

The third part of the lemma follows from the first two parts.

The proof of Theorem 3.1 relies on the following two lemmas.

**Lemma A.1.** *Let $\lambda_1 \leq \cdots \leq \lambda_I$ denote the eigenvalues of $H_{\mathcal{I}}$. The matrix $(\mathbf{I} - r\,H_{\mathcal{I}})$ is non-singular if and only if $\lambda_i \neq 1/r$, for every $i = 1, \ldots, I$. If $(\mathbf{I} - r\,H_{\mathcal{I}})$ is non-singular, then it is positive definite if and only if $\lambda_I < 1/r$.*

*Proof.* Because for all $l \in \mathcal{R}$ and for all $r > 0$, $[\mathbf{I} - r\,H_{\mathcal{I}} - l\mathbf{I}] = -r[H_{\mathcal{I}} - (1-l)/r\,\mathbf{I}]$, then the $I$ eigenvalues of $(\mathbf{I} - r\,H_{\mathcal{I}})$ are $1 - r\lambda_1 \geq \cdots \geq 1 - r\lambda_I$ and the statements in the lemma follow directly. □

**Lemma A.2.** *(i) $(X^TX - rX_{\mathcal{I}}X_{\mathcal{I}}^T)$ is singular if and only if $(\mathbf{I} - r\,H_{\mathcal{I}})$ is singular.*
*(ii) $(X^TX - rX_{\mathcal{I}}X_{\mathcal{I}}^T)$ is positive definite if and only if $(\mathbf{I} - r\,H_{\mathcal{I}})$ is positive definite.*

*Proof.* To prove the lemma we use a formula for matrix inversion given in [14]. For every square matrix $W$ and any conforming rectangular matrices $U$ and $V$, assuming that each of the stated inverses exists:

$$(W + U^TV)^{-1} = W^{-1} - W^{-1}U^T(\mathbf{I} + VW^{-1}U^T)^{-1}VW^{-1}. \tag{A.1}$$

By applying formula (A.1) to the matrices $W = X^TX$, $U = -rX_{\mathcal{I}}^T$ and $V = X_{\mathcal{I}}^T$, an expression for the inverse of $(X^TX - rX_{\mathcal{I}}X_{\mathcal{I}}^T)$, when it exists, is given by:

$$(X^TX - rX_{\mathcal{I}}X_{\mathcal{I}}^T)^{-1} =$$
$$= (X^TX)^{-1} + r(X^TX)^{-1}X_{\mathcal{I}}(\mathbf{I} - r\,H_{\mathcal{I}})^{-1}X_{\mathcal{I}}^T(X^TX)^{-1}. \tag{A.2}$$

On the other hand, if we substitute $W = \mathbf{I}$, $U = -r(X^TX)^{-1}X_{\mathcal{I}}$ and $V = X_{\mathcal{I}}$ into Equation (A.1), an expression for $(\mathbf{I} - r\,H_{\mathcal{I}})^{-1}$ is given by

$$(\mathbf{I} - r\,H_{\mathcal{I}})^{-1} = \mathbf{I} + rX_{\mathcal{I}}^T(X^TX - rX_{\mathcal{I}}X_{\mathcal{I}}^T)^{-1}X_{\mathcal{I}}. \tag{A.3}$$

Thus, we can use formula (A.2) to verify the *"if" part* of proposition ($i$) and formula (A.3) to verify the *"only if" part*. With regard to proposition ($ii$) of the lemma, observe that if $(\mathbf{I}-r\,H_{\mathcal{I}})^{-1}$ is positive definite, then $X_{\mathcal{I}}^T(\mathbf{I}-r\,H_{\mathcal{I}})^{-1}X_{\mathcal{I}}$ is positive semi-definite and Equation (A.2) shows that $(X^TX-rX_{\mathcal{I}}X_{\mathcal{I}}^T)^{-1}$ can be written as the sum of a positive definite matrix, $(X^TX)^{-1}$, and a positive semi-definite matrix. As such, it is positive definite and $(X^TX-rX_{\mathcal{I}}X_{\mathcal{I}}^T)$ must be positive definite as well. Looking at Equation (A.3) and arguing in a similar manner, the necessary condition in proposition ($ii$) may be proved. □

### *Proof of Theorem 3.1*

**Part ($i$)** The assumption that $\lambda_i \neq 1/r$ for all $i = 1,\dots,I$ implies that

$$\tilde{\boldsymbol{\theta}} = (X^TX - rX_{\mathcal{I}}X_{\mathcal{I}}^T)^{-1}(X^T\boldsymbol{y} - rX_{\mathcal{I}}\boldsymbol{y}_{\setminus\mathcal{I}}) \tag{A.4}$$

is well defined in view of formula (A.2) and Lemma A.1, and the posterior $r$th moment of $w_{\setminus\mathcal{I}}(s)$, $E(w_{\setminus\mathcal{I}}^r(s)|\boldsymbol{y})$, is proportional to

$$\begin{aligned}\int w_{\setminus\mathcal{I}}^r(s)q(s)\,ds = \int &(\sigma^2)^{-(n-rI)/2-\alpha-1}\times\\ &\times\exp\{-1/(2\sigma^2)[\boldsymbol{y}^T\boldsymbol{y} - r\boldsymbol{y}_{\mathcal{I}}^T\boldsymbol{y}_{\mathcal{I}} - \tilde{\boldsymbol{\theta}}^T(X^TX - rX_{\mathcal{I}}X_{\mathcal{I}}^T)\tilde{\boldsymbol{\theta}} + 2/\beta]\}\times\\ &\times\exp\{-1/(2\sigma^2)(\boldsymbol{\theta}-\tilde{\boldsymbol{\theta}})^T(X^TX - rX_{\mathcal{I}}X_{\mathcal{I}}^T)(\boldsymbol{\theta}-\tilde{\boldsymbol{\theta}})\}\pi_1(\boldsymbol{\theta})\,d\boldsymbol{\theta}\,d\sigma^2. \end{aligned}\tag{A.5}$$

If condition ($a$) holds, then, by Lemma A.2, $(X^TX-rX_{\mathcal{I}}X_{\mathcal{I}}^T)$ is positive definite, and

$$\begin{aligned}E(w_{\setminus\mathcal{I}}^r(s)|\boldsymbol{y}) \le \text{const}\times\int(\sigma^2)^{-(n-rI)/2-\alpha-1}\times\\ \times\exp\{-1/(2\sigma^2)[\boldsymbol{y}^T\boldsymbol{y} - r\boldsymbol{y}_{\mathcal{I}}^T\boldsymbol{y}_{\mathcal{I}} - \tilde{\boldsymbol{\theta}}^T(X^TX - rX_{\mathcal{I}}X_{\mathcal{I}}^T)\tilde{\boldsymbol{\theta}} + 2/\beta]\}\,d\sigma^2.\end{aligned}\tag{A.6}$$

Using the expression for $(X^TX - rX_{\mathcal{I}}X_{\mathcal{I}}^T)^{-1}$ given in Equation (A.2) and the property that $H_{\mathcal{I}}$ commutes with $(\mathbf{I}-H_{\mathcal{I}})$, we obtain:

$$\begin{aligned}\boldsymbol{y}^T\boldsymbol{y} - r\boldsymbol{y}_{\mathcal{I}}^T\boldsymbol{y}_{\mathcal{I}} &- \tilde{\boldsymbol{\theta}}^T(X^TX - rX_{\mathcal{I}}X_{\mathcal{I}}^T)\tilde{\boldsymbol{\theta}}\\ &= \boldsymbol{y}^T(\mathbf{I}-H)\boldsymbol{y} - r\,\boldsymbol{y}_{\mathcal{I}}^T[\mathbf{I} + r\,H_{\mathcal{I}} + r^2\,H_{\mathcal{I}}(\mathbf{I}-r\,H_{\mathcal{I}})^{-1}H_{\mathcal{I}}]\boldsymbol{y}_{\mathcal{I}}\\ &\quad+\ 2r\boldsymbol{y}_{\mathcal{I}}^T[\mathbf{I} + r\,H_{\mathcal{I}}(\mathbf{I}-r\,H_{\mathcal{I}})^{-1}]X_{\mathcal{I}}^T(X^TX)^{-1}X^T\boldsymbol{y}\\ &\quad-\ r\boldsymbol{y}^TX(X^TX)^{-1}X_{\mathcal{I}}(\mathbf{I}-r\,H_{\mathcal{I}})^{-1}X_{\mathcal{I}}^T(X^TX)^{-1}X^T\boldsymbol{y}\\ &\quad=\ \boldsymbol{y}^T(\mathbf{I}-H)\boldsymbol{y} - r\,\boldsymbol{e}_{\mathcal{I}}^T(\mathbf{I}-r\,H_{\mathcal{I}})^{-1}\boldsymbol{e}_{\mathcal{I}} = \text{RSS}_{\setminus\mathcal{I}}^*(r).\end{aligned}$$

Thus, the integrand in Equation (A.6) is proportional to an inverse gamma density if conditions ($b$) and ($c$) hold. Sufficiency of conditions ($a$) − ($c$) is proved. Suppose now that any of conditions ($a'$) or ($b'$) or ($c'$) holds. If ($b'$) is true then, as $\sigma^2 \to \infty$, the integrand in (A.5) goes to zero too slowly and it is not integrable. On the other hand, if $(c')$ holds, because quadratic forms are continuous and because $\pi_1$ has full support, then there exists a neighborhood $C_1$ of $\tilde{\boldsymbol{\theta}}$ having positive Lebesgue measure such that $\text{RSS}^*_{\backslash\mathcal{I}}(r) + 2/\beta + (\boldsymbol{\theta} - \tilde{\boldsymbol{\theta}})^T(X^TX - rX_{\mathcal{I}}X_{\mathcal{I}}^T)(\boldsymbol{\theta} - \tilde{\boldsymbol{\theta}}) < 0$. Also, when $(a')$ holds, because $(X^TX - rX_{\mathcal{I}}X_{\mathcal{I}}^T)$ is non-positive-definite, non-singular matrix, we can find a set $C_2$, depending on $\beta$ and $\text{RSS}^*_{\backslash\mathcal{I}}(r)$, with positive Lebesgue measure, such that $\text{RSS}^*_{\backslash\mathcal{I}}(r) + 2/\beta + (\boldsymbol{\theta} - \tilde{\boldsymbol{\theta}})^T(X^TX - rX_{\mathcal{I}}X_{\mathcal{I}}^T)(\boldsymbol{\theta} - \tilde{\boldsymbol{\theta}}) < 0$, for all $\boldsymbol{\theta} \in C_2$. Thus, under either of conditions $(a')$ or $(c')$, the integrand in (A.5) approaches infinity at an exponential rate as $\sigma^2 \to 0$ for every $\boldsymbol{\theta}$ belonging to a set with positive Lebesgue measure. It follows that $E(w^r_{\backslash\mathcal{I}}(s)|\boldsymbol{y}) = \infty$.

**Part *(ii)*** If the standard noninformative prior $\pi(\boldsymbol{\theta}, \sigma^2) \propto 1/\sigma^2$ is used, we can obtain an expression for $\int w^r_{\backslash\mathcal{I}}(s)q(s)\,ds$ by setting $\alpha = 0$ and $\pi_1(\boldsymbol{\theta}) \propto 1$ and letting $\beta$ tend to infinity in Equation (A.5). Then, if condition $(a)$ holds, we have

$$\int \exp\{-1/(2\sigma^2)(\boldsymbol{\theta} - \tilde{\boldsymbol{\theta}})^T(X^TX - rX_{\mathcal{I}}X_{\mathcal{I}}^T)(\boldsymbol{\theta} - \tilde{\boldsymbol{\theta}})\}\pi_1(\boldsymbol{\theta})\,d\boldsymbol{\theta} = \\ = \left(2\pi\sigma^2|X^TX - rX_{\mathcal{I}}X_{\mathcal{I}}^T|^{-1}\right)^{k/2},$$

where here $|\cdot|$ denotes the determinant of its argument and

$$E(w^r_{\backslash\mathcal{I}}(s)|\boldsymbol{y}) \propto \left(2\pi|X^TX - rX_{\mathcal{I}}X_{\mathcal{I}}^T|^{-1}\right)^{k/2} \times \\ \times \int (\sigma^2)^{-(n-rI-k)/2-1} \exp\{-1/(2\sigma^2)\text{RSS}^*_{\backslash\mathcal{I}}(r)\}\,d\sigma^2. \quad \text{(A.7)}$$

The integral on the right-hand side is finite if conditions $(b)$ and $(c)$ (as given in the statement of part $(ii)$) hold and sufficiency in part $(ii)$ is shown. The proof of the *"only if" part* proceeds as in part $(i)$.

### *Proof of Corollary 3.1*

Let $E_j(w^r_{\backslash\mathcal{I}}(\boldsymbol{\theta}, \sigma^2)|\boldsymbol{y})$ denote the posterior $r$th moment of the weight function when the prior distribution for $(\boldsymbol{\theta}, \sigma^2)$ is given by $\pi_{11}(\boldsymbol{\theta}) \times \pi_{j2}(\sigma^2)$, for $j = 0, 1$. If $\lambda_i \neq 1/r$ for all $i = 1, \ldots, I$, then, $E_j(w^r_{\backslash\mathcal{I}}(\boldsymbol{\theta}, \sigma^2)|\boldsymbol{y})$ is proportional to

$$\int (\sigma^2)^{-(n-rI)/2} \exp\{-1/(2\sigma^2)[\text{RSS}^*_{\backslash\mathcal{I}}(r) + \\ + (\boldsymbol{\theta} - \tilde{\boldsymbol{\theta}})^T(X^TX - rX_{\mathcal{I}}X_{\mathcal{I}}^T)(\boldsymbol{\theta} - \tilde{\boldsymbol{\theta}})]\}\pi_{11}(\boldsymbol{\theta})\pi_{j2}(\sigma^2)\,d\boldsymbol{\theta}\,d\sigma^2 \quad \text{(A.8)}$$

where $\tilde{\boldsymbol{\theta}}$ is (well) defined in Equation (A.4). As shown in the proof of Theorem 3.1, if $\lambda_I < 1/r$, then $0 < \exp\{-(1/(2\sigma^2)(\boldsymbol{\theta} - \tilde{\boldsymbol{\theta}})^T(X^TX - rX_{\mathcal{I}}X_{\mathcal{I}}^T)(\boldsymbol{\theta} - \tilde{\boldsymbol{\theta}})\} \leq 1$ so that

$$E_j(w^r_{\setminus\mathcal{I}}(\boldsymbol{\theta},\sigma^2)|\boldsymbol{y}) \le \text{const}\times$$
$$\times \int (\sigma^2)^{-(n-rI)/2} \exp\{-1/(2\sigma^2)\text{RSS}^*_{\setminus\mathcal{I}}(r)\}\pi_{j2}(\sigma^2)\, d\sigma^2, \qquad j=0,1. \quad \text{(A.9)}$$

Applying inequality (A.9) with $j = 1$ and using the assumption that $\text{RSS}^*_{\setminus\mathcal{I}}(r) > 0$, we have

$$E_j(w^r_{\setminus\mathcal{I}}(\boldsymbol{\theta},\sigma^2)|\boldsymbol{y}) \le \text{const} \times \int (\sigma^2)^{-(n-rI)/2}\pi_{12}(\sigma^2)\, d\sigma^2,$$

and the latter integral is finite by assumption. To prove the second part of the corollary, we first note that Theorem 3.1 implies that if $\lambda_I > 1/r$, then $E_0(w^r_{\setminus\mathcal{I}}(\boldsymbol{\theta},\sigma^2)|\boldsymbol{y}) = \infty$ for any $\pi_{11}$ and $\pi_{02} \in \mathcal{F}_2$, whereas, if $\text{RSS}^*_{\setminus\mathcal{I}}(r) < 0$, then $E_0(w^2_{\setminus\mathcal{I}}(\boldsymbol{\theta},\sigma^2)|\boldsymbol{y}) = \infty$ for any $\pi_{02}$ in $\mathcal{F}_2$ having $\beta > -2/\text{RSS}^*_{\setminus\mathcal{I}}(r)$. As we noted in the proof of Theorem 3.1, in both cases we can find a subset $C$ of $\mathcal{R}^k$ having positive Lebesgue measure such that, for any $\boldsymbol{\theta} \in C$, $E_0(w^r_{\setminus\mathcal{I}}(\boldsymbol{\theta},\sigma^2)|\boldsymbol{y})$ is infinite because the integral with respect to $\sigma^2$ does not exist in any neighborhood of zero. Because $\pi_{12}$ is thick-tailed with respect to $\mathcal{F}_2$, then, for every fixed $B > 0$, there exists a $\sigma_0^2$ such that $\pi_{12}(\sigma^2) > B\pi_{02}(\sigma^2)$ for any $\sigma^2 < \sigma_0^2$. Thus, by Lemma 2.1 $E_0(w^2_{\setminus\mathcal{I}}(\boldsymbol{\theta},\sigma^2)|\boldsymbol{y}) = \infty$ for some $\pi_{02}$ in $\mathcal{F}_2$ implies $E_1(w^2_{\setminus\mathcal{I}}(\boldsymbol{\theta},\sigma^2)|\boldsymbol{y}) = \infty$ as well.

### *Proof of Corollary 3.2*

If $\lambda_I < 1/r$, inequality (A.9) holds for both the prior $\pi_{11}(\boldsymbol{\theta}) \times \pi_{02}(\sigma^2)$ and $\pi_{11}(\boldsymbol{\theta}) \times \pi_{12}(\sigma^2)$. Furthermore, if $\pi_{02}(\sigma^2)$ is a prior distribution in $\mathcal{F}_2$ with $\alpha > -(n-rI)/2$ and with $\beta$ such that $\text{RSS}^*_{\setminus i}(2) > -2/\beta$, then, if $\lambda_I < 1/r$, $\int(\sigma^2)^{-(n-rI)/2}\exp\{-1/(2\sigma^2)\text{RSS}^*_{\setminus\mathcal{I}}(r)\}\pi_{02}(\sigma^2)\, d\sigma^2$ is finite. By assumption, for any fixed $b > 0$ there exists a $\delta > 0$ such that $\pi_{12}(\sigma^2) < b\,\pi_{02}(\sigma^2)$ for any $\sigma^2 < \delta$. Next, split the integral on the right hand side in Equation (A.9) into the two portions over $(0,\delta)$ and $[\delta,\infty)$. By Lemma 2.1, $\int_0^\delta(\sigma^2)^{-(n-rI)/2}\exp\{-1/(2\sigma^2)$ $\text{RSS}^*_{\setminus\mathcal{I}}(r)\}\pi_{02}(\sigma^2)\; d\sigma^2 < \infty$ implies $\int_0^\delta(\sigma^2)^{-(n-rI)/2}\exp\{-1/(2\sigma^2)\text{RSS}^*_{\setminus\mathcal{I}}(r)\}$ $\pi_{12}(\sigma^2)\; d\sigma^2 < \infty$. For the portion over $(\delta,\infty)$, it is enough to observe that $\int_\delta^\infty(\sigma^2)^{-(n-rI)/2}\exp\{-1/(2\sigma^2)\text{RSS}^*_{\setminus\mathcal{I}}(r)\}\pi_{12}(\sigma^2)\, d\sigma^2 < \text{const}\times\int_\delta^\infty(\sigma^2)^{-(n-rI)/2}$ $\pi_{12}(\sigma^2)\, d\sigma^2$, which is finite by assumption.

Assume now that $\pi_{11}(\boldsymbol{\theta})$ is thick-tailed with respect to $\mathcal{F}_1$ and that $\lambda_I > 1/r$. It follows from $\lambda_I > 1/r$ together with $\lambda_i \neq 1/r$ for all $i = 1,\ldots,I-1$ that $(X^TX - rX_{\mathcal{I}}X_{\mathcal{I}}^T)/\sigma^2$ is a non-positive-definite, non-singular matrix, $\forall\sigma^2 > 0$. Thus, there exists a sequence $\{\boldsymbol{\theta}_t^0\}$ with $||\boldsymbol{\theta}_t^0|| \to \infty$, as $t \to \infty$, and a vector $\boldsymbol{\epsilon} = (\epsilon_0,\ldots,\epsilon_{k-1})$, with $\epsilon_j > 0$ for all $j = 0,1,\ldots,k-1$ such that $\lim_{t\to\infty} 1/(2\sigma^2)(\boldsymbol{\theta}_t - \tilde{\boldsymbol{\theta}})^T(X^TX - rX_{\mathcal{I}}X_{\mathcal{I}}^T)(\boldsymbol{\theta}_t - \tilde{\boldsymbol{\theta}}) = -\infty$, for all sequences $\{\boldsymbol{\theta}_t\}$ such that $\boldsymbol{\theta}_t^0 - \boldsymbol{\epsilon} < \boldsymbol{\theta}_t < \boldsymbol{\theta}^0 + \boldsymbol{\epsilon}$. Keeping in mind that $\pi_{11}(\boldsymbol{\theta})$ is thick-tailed with respect to $\mathcal{F}_1$, then $\lim_{t\to\infty}\exp\{-1/(2\sigma^2)(\boldsymbol{\theta}_t - \tilde{\boldsymbol{\theta}})^T(X^TX - rX_{\mathcal{I}}X_{\mathcal{I}}^T)(\boldsymbol{\theta}_t - \tilde{\boldsymbol{\theta}})\}\pi_{11}(\boldsymbol{\theta}_t) = \infty$, for all sequences $\{\boldsymbol{\theta}_t\}$ such that $\boldsymbol{\theta}_t^0 - \boldsymbol{\epsilon} < \boldsymbol{\theta}_t < \boldsymbol{\theta}_t^0 + \boldsymbol{\epsilon}$. It follows from Equation (A.8) that $E_1(w^r_{\setminus\mathcal{I}}(\boldsymbol{\theta},\sigma^2)|\boldsymbol{y}) = \infty$.

***Proof of Example 3.1***

To avoid heavy algebra, we consider only the case $\theta_0 = \tilde{\theta}$, although the result is true for an arbitrary $\theta_0$. If $h_{ii} = 1/2 + 1/\sum_{j=1}^n x_j^2$, then $X^T X - 2\boldsymbol{x}_i \boldsymbol{x}_i^T = -2$. Some algebraic manipulations yield

$$E(w_{\setminus i}^2(\theta, \sigma^2)|\boldsymbol{y}) \propto \int_0^\infty (\sigma^2)^{-(n-2)/2} \exp\{-\mathrm{RSS}_{\setminus i}^*/(2\sigma^2)\} \times \\ \times \left[\int_0^\infty \exp\{x/\sigma^2 - x^2\} x^{-1/2}\, dx\right] \pi_{12}(\sigma^2)\, d\sigma^2$$

and for the interior integral with respect to $x$ the following bounds hold:

$$\int_0^1 \exp\{x/\sigma^2 - x^2\}\, dx \le \int_0^\infty \exp\{x/\sigma^2 - x^2\} x^{-1/2} dx \\ \le \exp\{1/\sigma^2\} \int_0^1 x^{-1/2}\, dx + \int_1^\infty \exp\{x/\sigma^2 - x^2\}\, dx. \quad \text{(A.10)}$$

Furthermore, $\int_a^b \exp\{x/\sigma^2 - x^2\}\, dx \propto \exp\{(2\sigma^2)^{-2}\}$, for all $-\infty \le a < b \le \infty$. This fact and the second inequality in (A.10) imply that if $\pi_{12} \propto \exp(-(\sigma^2)^{-2} - \sigma^2)$ then $E(w_{\setminus i}^2(\boldsymbol{\theta}, \sigma^2)|\boldsymbol{y}) < \text{const} \times \int_0^1 (\sigma^2)^{-(n-2)/2} \exp\{-(\sigma^2)^{-2} - \sigma^2 - (\mathrm{RSS}_{\setminus i}^*/2 - 1)/\sigma^2\}\, d\sigma^2 + \text{const} \times \int_0^1 (\sigma^2)^{-(n-2)/2} \exp\{-\frac{3}{4}(\sigma^2)^{-2} - \sigma^2 - \mathrm{RSS}_{\setminus i}^*/(2\sigma^2)\}\, d\sigma^2 < \infty$. On the other hand, if $\pi_{12}(\sigma^2) \propto \exp(-(\sigma^2)^{-3/2} - \sigma^2)$, then the first inequality in (A.10) yields

$$E(w_{\setminus i}^2(\boldsymbol{\theta}, \sigma^2)|\boldsymbol{y}) > \text{const} \times \int_0^\infty (\sigma^2)^{-(n-2)/2} \times \\ \times \exp\{-(\sigma^2)^{-3/2} - \sigma^2 - \mathrm{RSS}_{\setminus i}^*/(2\sigma^2) + (2\sigma^2)^{-2}\}\, d\sigma^2 = \infty.$$

***Proof of Theorem 4.1***

To simplify the notation, in this proof we will write $A(\kappa)$ for $A(\mathcal{I}, r, \kappa)$, $B(\kappa)$ for $B(\mathcal{I}, r, \kappa)$ and $C$ for $C(\mathcal{I}, r)$. Simple algebraic manipulations show that

$$E(w_{\setminus \mathcal{I}}^r(m, \sigma^2, \kappa)|\boldsymbol{v}) \propto \int w_{\setminus \mathcal{I}}^r(m, \sigma^2, \kappa) q(m, \sigma^2, \kappa)\, dm\, d\sigma^2\, d\kappa \\ = \int (\sigma^2)^{-(n-rI)/2-1} \exp\Big\{-\frac{1}{2\sigma^2}\Big[A(\kappa)m^2 - 2B(\kappa)m + C\Big]\Big\} \pi_2(\kappa)\, dm\, d\sigma^2\, d\kappa \quad \text{(A.11)}$$

$$= \int (\sigma^2)^{-(n-rI)/2-1} \exp\Big\{-\frac{1}{2\sigma^2}\Big[C - \frac{B^2(\kappa)}{A(\kappa)}\Big]\Big\} \times \quad \text{(A.12)} \\ \times \exp\Big\{-\frac{A(\kappa)}{2\sigma^2}\Big[m - \frac{B(\kappa)}{A(\kappa)}\Big]^2\Big\} \pi_2(\kappa)\, dm\, d\sigma^2\, d\kappa.$$

Suppose first that conditions $(a),(b)$ and $(c)$ are satisfied. It follows from $(a)$ that $A(\kappa)>0$ for almost all $\kappa$, so that $\exp\left\{-A(\kappa)/(2\sigma^2)\,[m-B(\kappa)/A(\kappa)]^2\right\}$ is proportional to the normal density with mean $B(\kappa)/A(\kappa)$ and variance $\sigma^2/A(\kappa)$. Then, denoting by $\Phi$ the standard normal cumulative distribution function, integral (A.12) reduces to

$$(2\pi)^{1/2}\int(\sigma^2)^{-(n-rI-1)/2-1}\exp\Big\{-\frac{1}{2\sigma^2}\Big[C-\frac{B^2(\kappa)}{A(\kappa)}\Big]\Big\} \tag{A.13}$$
$$\times\,\big[1-\Phi\big(-\frac{B(\kappa)}{(\sigma^2A(\kappa))^{1/2}}\big)\big]A^{-1/2}(\kappa)\pi_2(\kappa)\,d\sigma^2\,d\kappa$$
$$\le\int(2\pi)^{1/2}(\sigma^2)^{-(n-rI-1)/2-1}\exp\Big\{-\frac{1}{2\sigma^2}\Big[C-\frac{B^2(\kappa)}{A(\kappa)}\Big]\Big\} \tag{A.14}$$
$$\times\,A^{-1/2}(\kappa)\pi_2(\kappa)\,d\sigma^2\,d\kappa.$$

Under conditions $(b)$ and $(c)$, the integrand in (A.14) is proportional to an inverse gamma density for almost all $\kappa$ and integral (A.14) is proportional to

$$\int\Big[C-\frac{B^2(\kappa)}{A(\kappa)}\Big]^{-(n-rI-1)/2}A^{-1/2}(\kappa)\pi_2(\kappa)\,d\kappa \tag{A.15}$$

Moreover, condition $(c)$ implies that $[C-B^2(\kappa)/A(\kappa)]^{-(n-rI-1)/2}$ is a bounded (continuous) function of $\kappa$ on $\mathcal{N}^c$ so that if $\int A^{-1/2}(\kappa)\pi_2(\kappa)\,d\kappa<\infty$ then integral (A.15) is finite. Condition $(a)$ implies that $\sum_{i\in\mathcal{I}}c_i^2/\sum_{i=1}^n c_i^2 = = \lim_{\kappa\to\infty} l(\mathcal{I},\kappa)<1/r$ or, equivalently, that $\sum_{i\notin\mathcal{I}}c_i^2-(r-1)\sum_{i\in\mathcal{I}}c_i^2>0$ so that, as $\kappa$ tends to infinity, $A(\kappa)$ behaves like $1/\kappa^2$. Hence, the finiteness of $\int A^{-1/2}(\kappa)\pi_2(\kappa)\,d\kappa$ is guaranteed by $\int\kappa\pi_2(\kappa)\,d\kappa<\infty$. Sufficiency of conditions $(a),(b)$ and $(c)$ follows.

Assume now that conditions $(a),(b)$ and $(d)$ hold. Then $E(w^r_{\setminus\mathcal{I}}(m,\kappa,\sigma^2)|\boldsymbol{v})$ is still proportional to integral (A.13). We will prove that under conditions $(b)$ and $(d)$ the integral is finite. It follows from condition $(d)$ that $B(\kappa)<0$ almost surely and, for every fixed $\epsilon>0$, we can find a constant $M_1>0$ such that

$$1-\Phi\Big(-\frac{B(\kappa)}{(\sigma^2A(\kappa))^{1/2}}\Big)\le\frac{1+\epsilon}{\sqrt{2\pi}}\times\frac{(\sigma^2A(\kappa))^{1/2}}{|B(\kappa)|}\times\exp\Big\{-1/(2\sigma^2)\frac{B^2(\kappa)}{A(\kappa)}\Big\},$$

$\forall\sigma^2<(1/M_1^2)B^2(\kappa)/A(\kappa)$. Therefore, an upper bound for integral (A.13) is

$$(1+\epsilon)\int_0^\infty\int_0^{M(\kappa)}(\sigma^2)^{-(n-rI)/2}\exp\{-\frac{C}{2\sigma^2}\}|B(\kappa)|^{-1}\pi_2(\kappa)\,d\sigma^2\,d\kappa+$$
$$+\int_0^\infty\int_{M(\kappa)}^\infty(\sigma^2)^{-(n-rI-1)/2-1}\exp\Big\{-\frac{1}{2\sigma^2}\Big[C-\frac{B^2(\kappa)}{A(\kappa)}\Big]\Big\}A^{-1/2}(\kappa)\pi_2(\kappa)\,d\sigma^2\,d\kappa$$
$$:=I_1+I_2$$

where $M(\kappa):=B^2(\kappa)/[M_1^2A(\kappa)]$. With regard to integral $I_1$, observe that $I_1\le(1+\epsilon)\int_0^\infty\int_0^{M(\kappa)}M(\kappa)(\sigma^2)^{-(n-rI)/2-1}\exp\{-C/2\sigma^2\}|B(\kappa)|^{-1}\pi_2(\kappa)d\sigma^2d\kappa$.

Under conditions $(b)$ and $(d)$, $(\sigma^2)^{-(n-rI)/2-1}\exp\{-1/(2\sigma^2)C\}$ is proportional to an inverse gamma density so that $I_1 \leq M_2 \int_0^\infty M(\kappa)/|B(\kappa)|\pi_2(\kappa)\,d\kappa$ $= M_2/M_1^2 \int_0^\infty |B(\kappa)|/A(\kappa)\pi_2(\kappa)\,d\kappa$, for some constant $M_2 > 0$. Moreover, as $\kappa$ tends to $\infty$, it follows from condition $(d)$ that $|B(\kappa)|$ behaves like $1/\kappa$ and, as seen earlier, it follows from condition $(a)$ that $A(\kappa)$ behaves like $1/\kappa^2$. Hence, we conclude that $\int \kappa\pi(\kappa) < \infty$ implies $\int |B(\kappa)|/A(\kappa)\pi_2(\kappa)\,d\kappa < \infty$.
With regard to $I_2$, under condition $(b)$ we obtain:

$$I_2 \leq \int_0^\infty M(\kappa)^{-(n-rI-1)/2} \max\Big\{1, \exp\Big\{-\frac{C - B^2(\kappa)/A(\kappa)}{2M(\kappa)}\Big\}\Big\} A^{-1/2}(\kappa)\pi_2(\kappa)\;d\kappa.$$

Conditions $(a)$ and $(d)$ together yield $\sup_{\kappa\in\mathcal{N}^c} B^2(\kappa)/A(\kappa) < \infty$ and $\inf_{\kappa\in\mathcal{N}^c} M(\kappa) > 0$ and previous integral is finite if $\int A^{-1/2}(\kappa)\pi_2(\kappa)\,d\kappa < \infty$. Sufficiency of conditions $(a), (b), (d)$ follows.

Conversely, if $(e)$ holds, then the integrand in integral (A.11) approaches infinity at an exponential rate as $\sigma^2$ goes to zero, whereas if $n - rI \leq 0$, the integrand approaches zero too slowly as $\sigma^2$ goes to infinity. Both $(e)$ and $n-rI \leq 0$ imply that integral (A.11) is infinite. Actually, non integrability follows even if $0 < n-rI \leq 1$. To show this suppose that $A(\mathcal{I},r,\kappa)m^2-2B(\mathcal{I},r,\kappa)m+C(\mathcal{I},r) > 0$ for almost all $\kappa$ and that $n - rI > 0$. Thus, integral (A.11) is proportional to

$$\int \Big[A(\mathcal{I},r,\kappa)m^2 - 2B(\mathcal{I},r,\kappa)m + C(\mathcal{I},r)\Big]^{-(n-rI)/2} \pi_2(\kappa)\,dm\,d\kappa,$$

but the interior integral with respect to $m$ is infinite if $(n - rI) \leq 1$. Thus condition $(f)$ implies $E(w^r_{\setminus\mathcal{I}}(m,\kappa,\sigma^2)|\boldsymbol{v}) = \infty$.

The proof of Theorem 5.1 relies on the following lemma which relates a bound in terms of polar coordinates to the finiteness of the integral.

**Lemma A.3.** *Suppose that $f(\boldsymbol{\beta})$ is continuous in $\boldsymbol{\beta}$, $\boldsymbol{\beta} \in \mathcal{R}^k$, and that, for some $M < \infty$ and $b < 0$, $|f(\boldsymbol{\beta})| \leq \exp(b\,||\boldsymbol{\beta}||)$ for all $\boldsymbol{\beta}$ such that $||\boldsymbol{\beta}|| \geq M$. Then $\int_{\mathcal{R}^k} |f(\boldsymbol{\beta})|d\boldsymbol{\beta} < \infty$.*

*Proof.* Split the integral into two portions. For $\boldsymbol{\beta}$ such that $||\boldsymbol{\beta}|| \leq M$, we have the integral of a continuous function over a compact set. This integral is finite. The integral over the remaining portion of the space is also finite:

$$\int_{||\boldsymbol{\beta}||>M} |f(\boldsymbol{\beta})|d\boldsymbol{\beta} \leq \int_{||\boldsymbol{\beta}||>M} \exp(b\;||\boldsymbol{\beta}||)d\boldsymbol{\beta} = \int_M^\infty c_k r^{k-1}\exp(br)\,dr < \infty,$$

where $c_k r^{k-1}$ is the surface area of the $k$ dimensional sphere of radius $r$. □

### *Proof of Theorem 5.1*

The expected $r$th moment of the case-deleted weight function can be written as an integral against the prior times the likelihood:

$$\int w^r_{\setminus\mathcal{I}}(\boldsymbol{\beta})\pi(\boldsymbol{\beta})f(y|x,\boldsymbol{\beta})d\boldsymbol{\beta} =$$
$$= \int \prod_{i\in\mathcal{I}} \frac{(1+\exp\{\boldsymbol{\beta}^T\boldsymbol{x}_i\})^{r-1}}{\exp\{(r-1)\boldsymbol{\beta}^T\boldsymbol{x}_i y_i\}} \exp(-\epsilon|\boldsymbol{\beta}^T|\mathbf{1}) \prod_{i\notin\mathcal{I}} \frac{\exp\{\boldsymbol{\beta}^T\boldsymbol{x}_i y_i\}}{1+\exp\{\boldsymbol{\beta}^T\boldsymbol{x}_i\}} d\boldsymbol{\beta}.$$

In order to apply Lemma A.3, we consider a ray emanating from the origin in an arbitrary direction, specified by a particular $\boldsymbol{\beta}$ under the constraint that $|\boldsymbol{\beta}^T|\mathbf{1} = 1$. In this fixed direction, the rate of decay (or increase) of the tail is determined by the maximum contribution, either 1 or $\exp\{\boldsymbol{\beta}^T\boldsymbol{x}_i\}$, from each term of the form $1+\exp\{\boldsymbol{\beta}^T\boldsymbol{x}_i\}$ in the products above. Collecting terms, we have that the rate of decay is governed by

$$\exp\Bigg(\sum_{i\notin\mathcal{I}} \boldsymbol{\beta}^T\boldsymbol{x}_i y_i - (r-1)\sum_{i\in\mathcal{I}} \boldsymbol{\beta}^T\boldsymbol{x}_i y_i - \sum_{i\notin\mathcal{I}} \max(0,\boldsymbol{\beta}^T\boldsymbol{x}_i) +$$
$$+ (r-1)\sum_{i\in\mathcal{I}} \max(0,\boldsymbol{\beta}^T\boldsymbol{x}_i) - \epsilon|\boldsymbol{\beta}^T|\mathbf{1}\Bigg) = \exp(h(\boldsymbol{\beta},r,\epsilon))$$

We consider the expression above, and note that we can obtain an (decreasing) exponential bound on the tail whenever the term inside the exponential is negative. If the corresponding expression is negative for every direction, we can construct a uniform bound which satisfies the assumption of the lemma which, in turn, allows us to conclude that the $r$th moment of the case-deleted weight function is finite.

The infinite $r$th moment case involves a positive value for some direction specified by $\boldsymbol{\beta}$. In this event, since $h(\boldsymbol{\beta},r,\epsilon)$ is continuous in $\boldsymbol{\beta}$, we conclude that there is a neighborhood of directions in which the integral along a ray is infinite. Thus, the integral is infinite, and so is the $r$th moment of the case-deleted weight function.